\documentclass{aastex631}
\usepackage{amsmath}
\usepackage{xurl}
\usepackage{graphicx}
\usepackage{xcolor}
\hypersetup{colorlinks=true, linkcolor=blue, citecolor=blue, urlcolor=blue}
\newcommand{\Pal}{Palomar~5}

\newcommand{\phione}{\phi_1}
\newcommand{\phitwo}{\phi_2}
\newcommand{\qphiz}{q_z}
\newcommand{\kms}{\mathrm{km\,s^{-1}}}
\newcommand{\masyr}{\mathrm{mas\,yr^{-1}}}
\newcommand{\msun}{\mathrm{M_\odot}}
\graphicspath{{figures/}}

\shorttitle{The Palomar 5 Stream in Legacy Surveys DR10}
\shortauthors{Xiao et al.}

\begin{document}

\title{The Palomar 5 Stellar Stream in DESI Legacy Surveys DR10: Wide-field Morphology and Constraints on the Inner Halo Potential Shape}

\author[0009-0004-2243-8289]{Yun-Ao Xiao}
\affiliation{National Astronomical Observatories, Chinese Academy of Sciences, Beijing 100101, People's Republic of China}
\affiliation{University of Chinese Academy of Sciences, Beijing 100049, People's Republic of China}

\author[0000-0002-6684-3997]{Hu Zou}
\affiliation{National Astronomical Observatories, Chinese Academy of Sciences, Beijing 100101, People's Republic of China}
\affiliation{University of Chinese Academy of Sciences, Beijing 100049, People's Republic of China}

\author{Lu Feng}
\affiliation{National Astronomical Observatories, Chinese Academy of Sciences, Beijing 100101, People's Republic of China}

\author[0009-0005-5203-3802]{Jinfu Gou}
\affiliation{National Astronomical Observatories, Chinese Academy of Sciences, Beijing 100101, People's Republic of China}
\affiliation{University of Chinese Academy of Sciences, Beijing 100049, People's Republic of China}

\author[0000-0001-9457-0589]{Wei-Jian Guo}
\affiliation{National Astronomical Observatories, Chinese Academy of Sciences, Beijing 100101, People's Republic of China}

\author[0000-0002-0656-075X]{Niu Li}
\affiliation{National Astronomical Observatories, Chinese Academy of Sciences, Beijing 100101, People's Republic of China}

\author[0000-0002-0096-3523]{Wenxiong Li}
\affiliation{National Astronomical Observatories, Chinese Academy of Sciences, Beijing 100101, People's Republic of China}

\author[0009-0007-8264-8343]{Shufei Liu}
\affiliation{National Astronomical Observatories, Chinese Academy of Sciences, Beijing 100101, People's Republic of China}
\affiliation{University of Chinese Academy of Sciences, Beijing 100049, People's Republic of China}

\author[0009-0000-8668-6717]{Haoming Song}
\affiliation{National Astronomical Observatories, Chinese Academy of Sciences, Beijing 100101, People's Republic of China}
\affiliation{University of Chinese Academy of Sciences, Beijing 100049, People's Republic of China}

\author[0000-0003-2132-0727]{Jipeng Sui}
\affiliation{National Astronomical Observatories, Chinese Academy of Sciences, Beijing 100101, People's Republic of China}
\affiliation{University of Chinese Academy of Sciences, Beijing 100049, People's Republic of China}

\author[0009-0006-0370-6828]{Xi Tan}
\affiliation{National Astronomical Observatories, Chinese Academy of Sciences, Beijing 100101, People's Republic of China}
\affiliation{University of Chinese Academy of Sciences, Beijing 100049, People's Republic of China}

\author[0000-0002-7048-0930]{Jingyi Zhang}
\affiliation{National Astronomical Observatories, Chinese Academy of Sciences, Beijing 100101, People's Republic of China}

\correspondingauthor{Hu Zou}
\email{zouhu@nao.cas.cn}

\begin{abstract}
The \Pal\ stellar stream probes the gravitational potential of the inner Milky Way, where the disk, dark halo, and time-dependent perturbations all contribute substantially to the force field. Using the DESI Legacy Imaging Surveys Data Release 10 (DR10), we map the wide-field morphology of the stream over $-25^\circ<\phione<20^\circ$ and $-5^\circ<\phitwo<10^\circ$. Our analysis combines a cluster-calibrated PARSEC color--magnitude diagram filter, a continuous main-sequence turnoff plus RR Lyrae distance relation, and a field star response tied to local imaging depth. After applying the response correction, we recover the stream clearly over $-15.5^\circ\le\phione\le7.0^\circ$ and reveal no coherent low-surface-brightness extension beyond this range within our DR10 search footprint. Mock stream fits to the measured centroid and width yield an NFW potential flattening parameter $\qphiz=0.82^{+0.13}_{-0.11}$, assuming a static Hernquist bulge, a Miyamoto--Nagai disk, and a potential-flattened NFW halo with fixed $r_s=20\,{\rm kpc}$ and fixed $V_c(R_0)=232\,\kms$ at the Solar radius. The posterior favors an oblate effective inner halo potential, with a minor dependence on the assumed progenitor mass. The DR10 stream morphology therefore provides a homogeneous observational reference well suited for future joint analyses that incorporate density information, proper motions, line-of-sight velocities, and time-dependent perturbations.
\end{abstract}


\section{Introduction}
\label{sec:intro}

Stellar streams serve as long-baseline dynamical probes of the Milky Way. When stars are stripped from a globular cluster or dwarf galaxy, they initially follow orbits close to that of the progenitor. Small differences in energy and angular momentum subsequently stretch the debris into a coherent structure in angle and frequency space \citep[e.g.,][]{HelmiWhite1999}. The resulting stream track, distance gradient, proper motion field, radial velocities, cross-stream width, and density structure collectively encode information about the Galactic potential and the time-dependent perturbations that shape the debris. This framework has been applied to massive dwarf galaxy streams such as Sagittarius, which probe the outer halo potential \citep{Ibata2001Sgr,Majewski2003,LawMajewski2010}, and to thin globular cluster streams such as GD-1, which are sensitive to the local acceleration field and dark substructure \citep{Koposov2010,Bonaca2019GD1,NibauerBonaca2025}. The rapid growth of wide-field survey samples (e.g., DES and Gaia) has since shifted the emphasis from individual stream fits toward global mass modeling of the halo \citep{Shipp2018,Ibata2024}. In this regime, non-static effects such as the influence of the Large Magellanic Cloud (LMC) are no longer negligible, as demonstrated by the Orphan stream \citep{Erkal2019Orphan}.

The \Pal\ stream offers a complementary but distinct probe: it lies closer to the Galactic center, where the disk, bar, and inner halo jointly shape the potential. SDSS star counts first revealed long tidal tails around this sparse, low-mass cluster \citep{Odenkirchen2001,Rockosi2002}, and later matched-filter maps traced the debris over more than $20^\circ$ on the sky \citep{Odenkirchen2003,GrillmairDionatos2006}. The tails exhibit clear density variations and a strong leading/trailing asymmetry. These features have inspired interpretations ranging from epicyclic structure and smooth halo evolution \citep{MastrobuonoBattisti2012,Kupper2015} to perturbations by dark subhalos, baryonic structures, and the Galactic bar \citep{Carlberg2012Pal5,Pearson2017,Erkal2017}. However, the reality of some density substructures remains debated, as spatially varying survey depth, extinction, and foreground contamination can imprint artificial clumps in matched-filter maps \citep{Thomas2016}. A robust measurement of \Pal\ therefore requires both a realistic stream model and a detailed survey response model to be treated jointly.

\Pal\ has also served as a direct probe of the inner Milky Way potential. Its morphology is sensitive to halo flattening because changes in the vertical and radial force components modify both the projected curvature of the stream and the angular separation between the leading and trailing tails \citep{Pearson2015}. Joint modeling of \Pal\ and GD-1 has already been used to constrain the shape of the dark halo inside roughly 20 kpc \citep{Bovy2016}, and \Pal\ remains one of the few globular cluster streams with a known progenitor, well-characterized photometric density maps, multi-tracer distance estimates, and a growing kinematic sample. RR Lyrae stars provide precise distance and kinematic constraints along the stream \citep{PriceWhelan2019}, while spectroscopy combined with Gaia astrometry yields high-confidence member samples and independent evidence for leading arm fanning \citep{Kuzma2022}. Searches for red giant members using APOGEE and Gaia further contribute chemical abundance constraints, while also underscoring the difficulty of detecting additional giants in the most diffuse regions of the tails \citep{Sheffield2025}.

Deep CFHT/MegaCam $g,r$ imaging allowed \citet{Ibata2016Pal5Photometry} to map the \Pal\ stream track and measure its relative line-of-sight distance variation. \citet{Ibata2017Pal5Mass} combined this photometry with spectroscopy to identify stream members and constrain the remnant structure and the mass in the cluster and tails. Using the same CFHT catalog, \citet{Erkal2017} jointly recovered the stream density, centroid track, and width with an adaptive probabilistic model. \citet{Bonaca2020} further characterized the projected \Pal\ morphology. Their analysis, based on DECam Legacy Survey (DECaLS) imaging, revealed substantial variations in width, density, and orientation, including a broad, low-surface-brightness leading component. In their fiducial model, a rotating Galactic bar reproduced part of the observed width and density structure, although the trailing arm track remained unexplained. Subsequent work has expanded the theoretical landscape. A large retained population of stellar-mass black holes can affect both the present-day cluster structure and the tidal debris \citep{Gieles2021}. A tilting Galactic disk can produce fanning and track-width patterns that could be misinterpreted in a static potential model \citep{Nibauer2024}. Encounters with known globular clusters can create gaps in the \Pal\ tails \citep{Ferrone2025}. Consequently, flexible stream inference methods are increasingly moving toward simultaneous modeling of morphology, kinematics, and time-dependent perturbations \citep{Nibauer2025StreamSculptor,Hilmi2024}.

The wide footprint of the DESI Legacy Imaging Surveys Data Release 10 (hereafter DR10), including extensive coverage away from the stream, enables us to revisit the low-contrast outskirts of Palomar 5 using a uniform photometric selection and field star response model across the entire search region. It also provides a consistently defined centroid and width measurement for testing the inner halo force field, provided that the potential convention and the angular range used in the fit are specified explicitly. In this work, we use DR10 photometry to construct a distance-rescreened Palomar 5 sample, measure the local centroid and width of the stream, and fit these measurements with mock streams in a smooth bulge+disk+halo potential.

The structure of the paper is as follows. Section~\ref{sec:data} describes the DR10 catalog, the color--color and CMD-based member selection, and the adopted isochrone calibration. Section~\ref{sec:morph} presents the distance modulus (DM) model, the exposure and field star background response, and the local stream morphology measurement. Section~\ref{sec:model} defines the mock stream construction, the track-width likelihood, and the halo flattening results. Section~\ref{sec:discussion} relates these results to previous \Pal\ and halo shape studies, and Section~\ref{sec:summary} summarizes our conclusions.

\section{Data and Photometric Member Selection}
\label{sec:data}

We use point-source photometry from DR10 in the $g,r,z$ bands. The Legacy Surveys comprise three complementary imaging programs: the Dark Energy Camera Legacy Survey (DECaLS) using the Blanco 4 m telescope \citep{Blum2016DECaLS,Dey2019}, the Beijing-Arizona Sky Survey (BASS) using the Bok 2.3 m telescope \citep{Zou2017BASS,Zou2019BASSDR3}, and the Mayall $z$-band Legacy Survey (MzLS) using the Mayall 4 m telescope \citep{Silva2016MzLS,Dey2019}. These surveys were designed to provide deep and uniform $g,r,z$ imaging for DESI target selection. DR10 further expands the Legacy Surveys by incorporating additional public DECam imaging over a footprint exceeding $2\times10^4\,\mathrm{deg^2}$\footnote{\url{https://www.legacysurvey.org/dr10/description/}}. In the southern DECaLS footprint where Palomar 5 is located, the characteristic $5\sigma$ point-source depths are approximately $g=24.7$, $r=23.9$, and $z=23.0$. The catalog depth map in the \Pal\ region shows local variations around the nominal southern depth values, with the best-covered portions of the stream footprint reaching approximately one magnitude deeper in $g$ band.

The initial catalog is preselected in a broad box around \Pal. The RA--Dec coordinates are transformed into the \Pal\ stream coordinate system implemented in \texttt{gala} \citep{PriceWhelan2017Gala}. Appendix~\ref{app:coords} gives the rotation matrix for this transformation. We mask the neighboring globular cluster M5 and retain sources in the stream coordinate window (see Figure~\ref{fig:exposure}):
\begin{equation}
-25^\circ<\phione<20^\circ,\qquad -5^\circ<\phitwo<10^\circ.
\end{equation}

\begin{figure*}[t]
\centering
\begin{minipage}{0.485\textwidth}
\centering
\includegraphics[width=\linewidth]{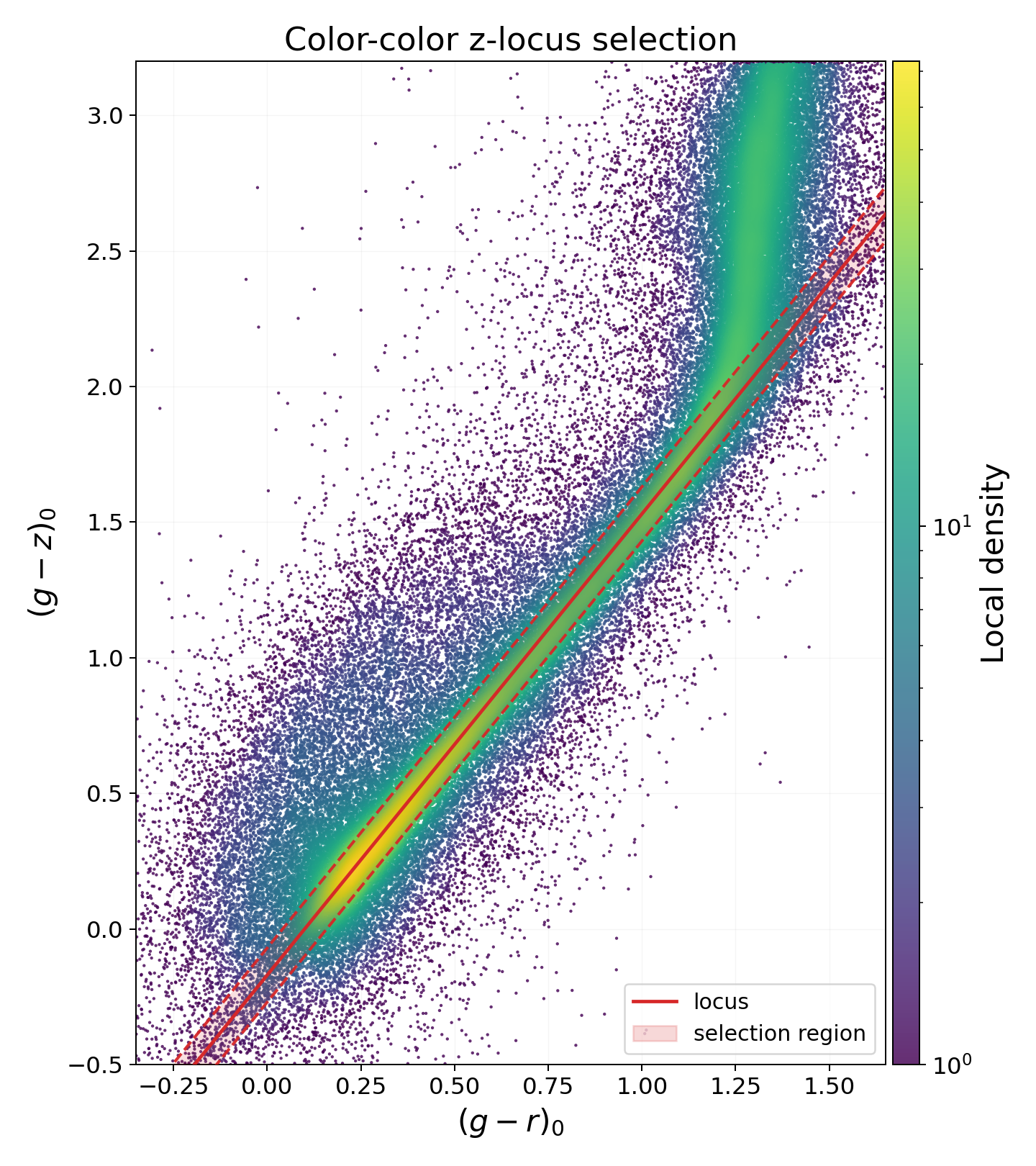}
\end{minipage}\hfill
\begin{minipage}{0.485\textwidth}
\centering
\includegraphics[width=\linewidth]{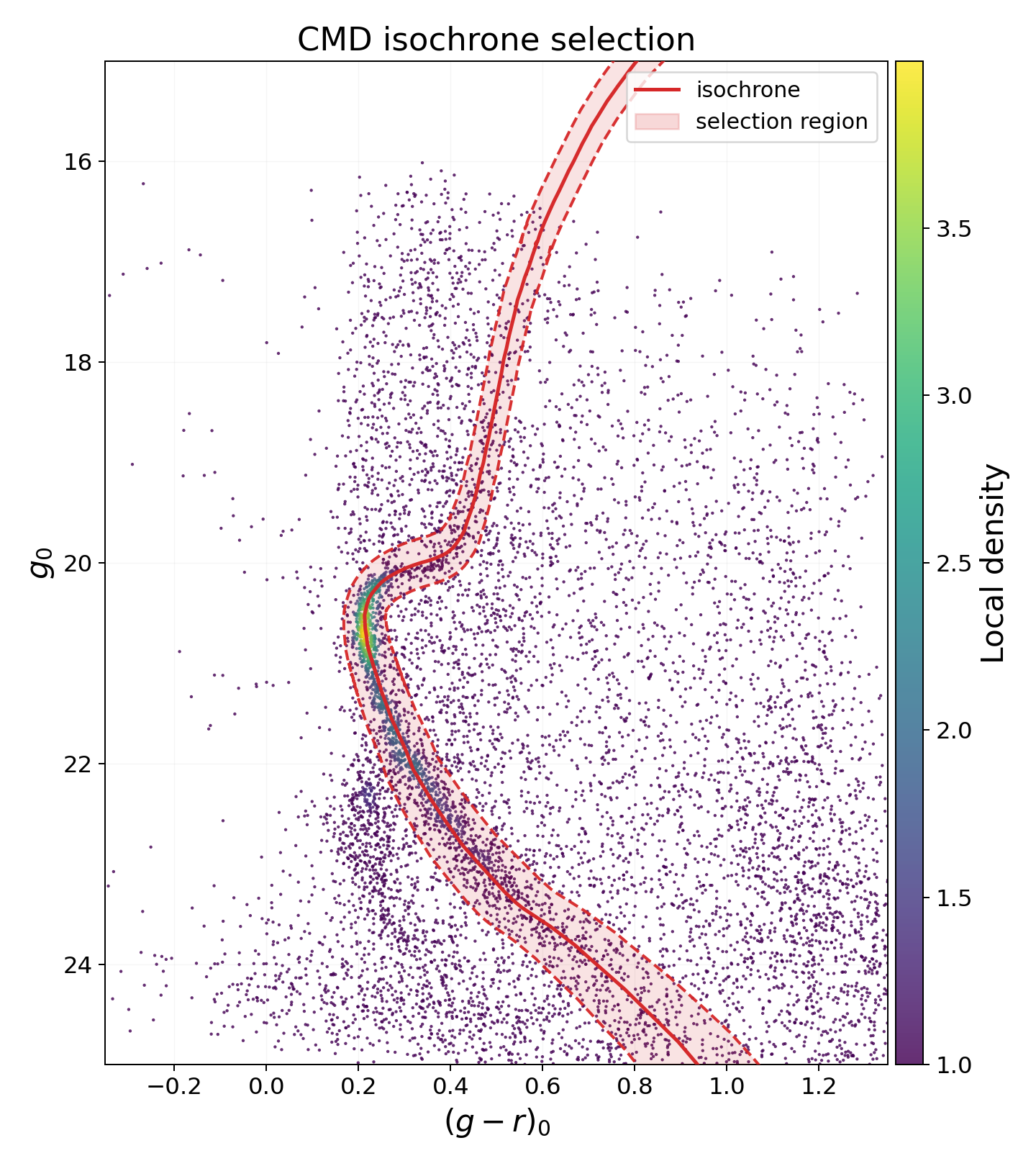}
\end{minipage}
\caption{Photometric selection diagnostics. Left: dereddened color--color diagram and the adopted filter in $(g-r)_0$ and $(g-z)_0$. Right: CMD of stars within $0.5^\circ$ of the \Pal\ center, with the empirically aligned PARSEC isochrone and the CMD envelope used for member selection. The two panels illustrate the color--color and CMD components of the selection function.}
\label{fig:memberselection}
\end{figure*}

Further member selection begins with broad photometric criteria: $15\le g_0<25$ and $-0.35\le (g-r)_0\le1.65$. We retain point sources with \texttt{TYPE=PSF}, require finite $g,r,z$ photometry, and reject sources affected by the Legacy Surveys mask bits in the relevant bands. The photometry is corrected for foreground extinction using the \citet{Schlegel1998} dust map. We then apply an empirical color--color filter in the dereddened $(g-r)_0$--$(g-z)_0$ plane,
\begin{equation}
(g-z)_0 = 1.7\,(g-r)_0 - 0.17,
\end{equation}
with a tolerance of 0.10 mag. This linear color locus follows the color--color selection used by \citet{Bonaca2020}. The left panel of Figure~\ref{fig:memberselection} shows the retained locus in color--color space. The filter suppresses broad disk stars and galaxy contamination while retaining the old, metal-poor main-sequence and turnoff locus of \Pal.

The CMD filter is anchored to stars within $0.5^\circ$ of the \Pal\ center. We generate the reference isochrone from PARSEC evolutionary tracks \citep{Bressan2012} using the PARSEC CMD interface\footnote{\url{https://stev.oapd.inaf.it/cgi-bin/cmd}} in the DECam $g,r,z$ system. The stellar population parameters, age $=11.508\,{\rm Gyr}$ and $[\mathrm{Fe/H}]=-1.798$, are taken from \citet{Xu2021Pal5}. We empirically align the isochrone with the observed cluster sequence, obtaining a reference CMD placement of ${\rm DM}_{{\rm CMD},0}=16.685$ mag. The CMD envelope is defined by the offset from the isochrone. The maximum acceptable offset to the isochrone is modeled as an intrinsic dispersion plus the photometric uncertainty,
\begin{equation}
w(g_0)=w_{\rm int}+(\sigma^2_{g-r}(g_0)+\sigma_g^2(g_0))^{1/2},
\end{equation}
where $\sigma_{g-r}=(\sigma_g^2+\sigma_r^2)^{1/2}$, and $\sigma_g$ and $\sigma_r$ are estimated via polynomial fits to the photometric errors as a function of magnitude. The adopted intrinsic term is $w_{\rm int}=0.04$ mag. Stars are retained if their offset distance from the isochrone lies within this magnitude-dependent envelope. The right panel of Figure~\ref{fig:memberselection} shows stars within $0.5^\circ$ of the cluster center, the aligned isochrone, and the adopted CMD envelope.

For the first-pass stream selection, we initialize the arm-dependent CMD shifts using approximate distances of $D=23$ kpc on the trailing side and $D=19$ kpc on the leading side before applying the cluster-based alignment. These values are used only to keep the initial filter close to the known distance gradient of \Pal, for which the leading arm is closer and the trailing arm is farther than the cluster \citep{PriceWhelan2019}. After constructing the continuous DM model in Section~\ref{subsec:distance}, we apply the updated DM to each star and repeat the color--color and CMD selection. The resulting sample is then used for the projected morphology analysis.

\section{Distance-rescreened Stream Morphology}
\label{sec:morph}

\subsection{Distance model for the CMD filter}
\label{subsec:distance}

The distance of \Pal\ varies measurably along the stream. A fixed apparent-magnitude CMD filter would follow the cluster sequence near the progenitor but shift relative to the leading and trailing debris at large $|\phione|$. We therefore construct a continuous DM curve and incorporate it into the photometric selection function.

\begin{figure*}[t]
\centering
\includegraphics[width=0.98\textwidth]{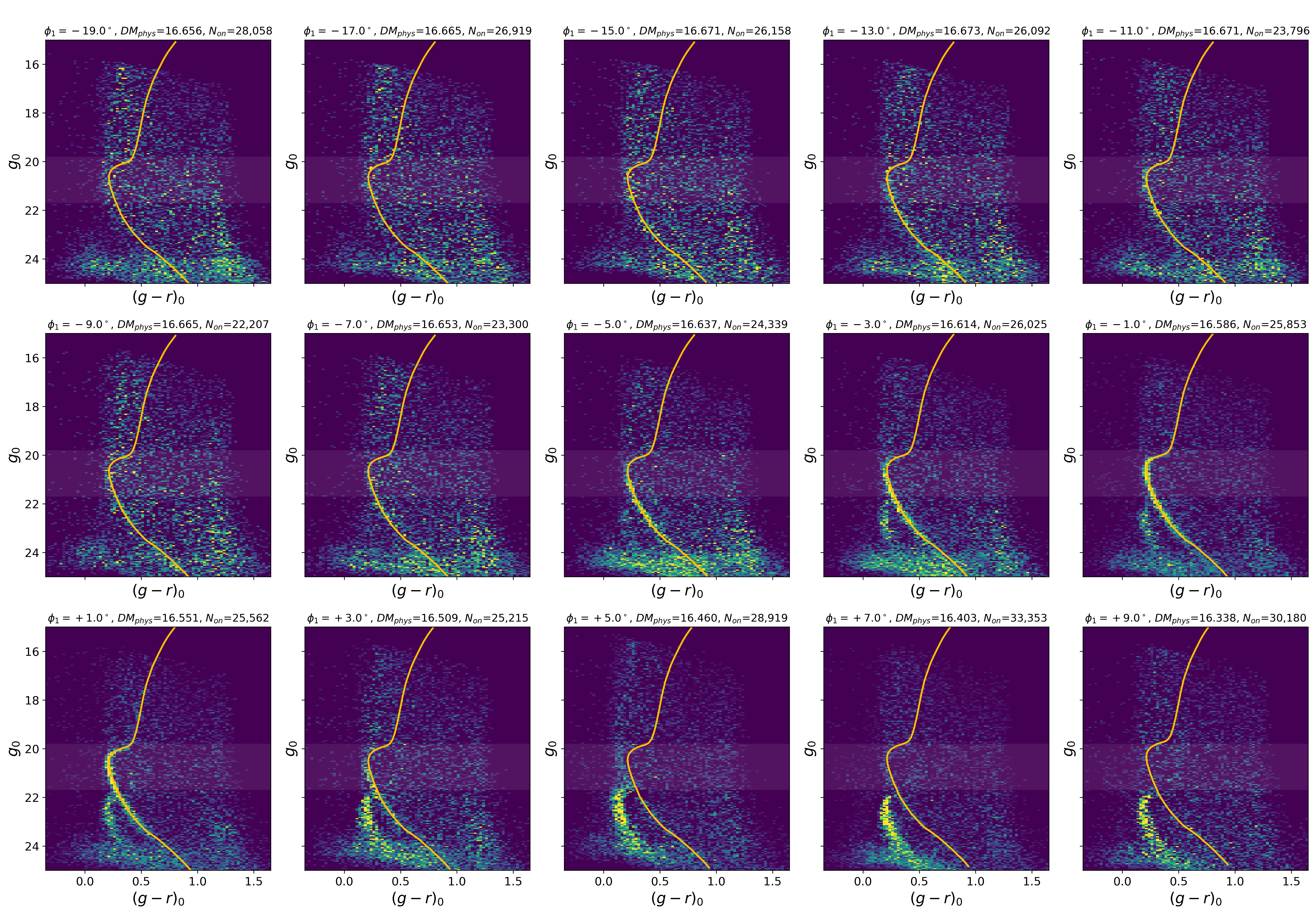}
\caption{Background-subtracted CMD diagnostics for the MSTO DM scan. Each panel corresponds to a different $\phione$ segment and shows the on-stream CMD after subtraction of the scaled local off-stream/background CMD, with the shifted isochrone overplotted. The translucent horizontal band marks the $19.8<g_0<21.7$ MSTO magnitude window used to evaluate the photometric distance score. The persistence of the turnoff and upper main sequence after local background subtraction motivates the continuous $\text{DM}(\phione)$ curve used for the final CMD rescreening.}
\label{fig:segmentcmds}
\end{figure*}

Starting from the first-pass member sample, we define overlapping windows centered at $\phione=-19^\circ,-17^\circ,\ldots,7^\circ,9^\circ$, with half-width $1.5^\circ$. The first-pass projected density fit gives the local stream center $\mu_0(\phione)$. Stars with $|\phitwo-\mu_0|<0.4^\circ$ define the on-stream CMD, while $0.8^\circ<|\phitwo-\mu_0|<1.6^\circ$ defines the off-stream CMD. We scan the isochrone placement ${\rm DM}_{\rm CMD}$ in steps of 0.01 mag over $\pm0.45$ mag around the initial value, using the MSTO range $19.8<g_0<21.7$. Figure~\ref{fig:segmentcmds} shows the background-subtracted CMDs.

At each trial ${\rm DM}_{\rm CMD}$, the cluster-aligned isochrone is shifted by ${\rm DM}_{\rm CMD}-{\rm DM}_{{\rm CMD},0}$. For the $j$th star, the color residual is

\begin{equation}
\Delta c_j({\rm DM}_{\rm CMD})=(g-r)_{0,j}-(g-r)_{\rm iso}(g_{0,j}\mid{\rm DM}_{\rm CMD}).
\end{equation}

The corresponding unnormalized Gaussian CMD weight is

\begin{equation}
W_{{\rm CMD},j}({\rm DM}_{\rm CMD})=\exp\left[-\frac{1}{2}\left(\frac{\Delta c_j({\rm DM}_{\rm CMD})}{w(g_{0,j})}\right)^2\right],
\label{eq:cmdweight}
\end{equation}

where $w(g_{0,j})$ is the magnitude-dependent CMD width defined in Section~\ref{sec:data}. The matched-filter score is then

\begin{equation}
S({\rm DM}_{\rm CMD})=\sum_{j\in{\rm on}}W_{{\rm CMD},j}({\rm DM}_{\rm CMD})-\alpha\sum_{j\in{\rm off}}W_{{\rm CMD},j}({\rm DM}_{\rm CMD}).
\label{eq:dmscore}
\end{equation}

The ratio of the on-stream and off-stream aperture widths gives $\alpha=0.5$. The best-fitting CMD modulus, ${\rm DM}_{{\rm CMD},{\rm best}}$, maximizes $S$.

To characterize the width of the distance-modulus peak, we fit the dominant maximum of the score curve with

\begin{equation}
S({\rm DM}_{\rm CMD})=S_0+A_{\rm DM}\exp\left[-\frac{({\rm DM}_{\rm CMD}-{\rm DM}_{{\rm CMD},{\rm best}})^2}{2\sigma_{\rm DM}^2}\right],
\label{eq:dmscoregauss}
\end{equation}

where $S_0$ is a local score baseline and $A_{\rm DM}$ is the peak amplitude. The quoted $\sigma_{\rm DM}$ is the fitted Gaussian width of the score peak. It characterizes the distance resolution of the MSTO matched-filter scan.

Before fitting ${\rm DM}(\phione)$, we calibrate all MSTO anchors to the RR Lyrae distance $D_0=20.6\pm0.2\,{\rm kpc}$ \citep{PriceWhelan2019}, corresponding to ${\rm DM}_0=16.569$ mag. The offset from the cluster CMD reference is $\delta_{\rm CMD}={\rm DM}_{{\rm CMD},0}-{\rm DM}_0=0.116$ mag, so we apply the same constant correction to every MSTO anchor:

\[
{\rm DM}_{{\rm phys},i}={\rm DM}_{{\rm CMD},i}-\delta_{\rm CMD}.
\]

Table~\ref{tab:mstoanchors} lists the calibrated values.

\begin{table}[t]
\centering
\caption{RR Lyrae-calibrated MSTO DM anchors from the background-subtracted CMD scans. The columns give $\phione$, the calibrated best-fitting DM, and the fitted Gaussian width $\sigma_{\rm DM}$ of the dominant score peak.}
\label{tab:mstoanchors}
\begin{tabular}{rrr}
\hline\hline
$\phione$ (deg) & $\text{DM}$ (mag) & $\sigma_{\text{DM}}$ (mag)\\
\hline
-19.0 & 16.613 & 0.120 \\
-17.0 & 16.483 & 0.130 \\
-15.0 & 16.553 & 0.143 \\
-13.0 & 16.893 & 0.161 \\
-11.0 & 16.673 & 0.131 \\
-9.0 & 16.763 & 0.221 \\
-7.0 & 16.743 & 0.120 \\
-5.0 & 16.613 & 0.127 \\
-3.0 & 16.593 & 0.144 \\
-1.0 & 16.603 & 0.146 \\
1.0 & 16.578 & 0.227 \\
3.0 & 16.398 & 0.147 \\
5.0 & 16.578 & 0.250 \\
7.0 & 16.578 & 0.250 \\
9.0 & 16.438 & 0.188 \\
\hline
\end{tabular}
\end{table}

\begin{figure*}[t]
\centering
\includegraphics[width=0.85\textwidth]{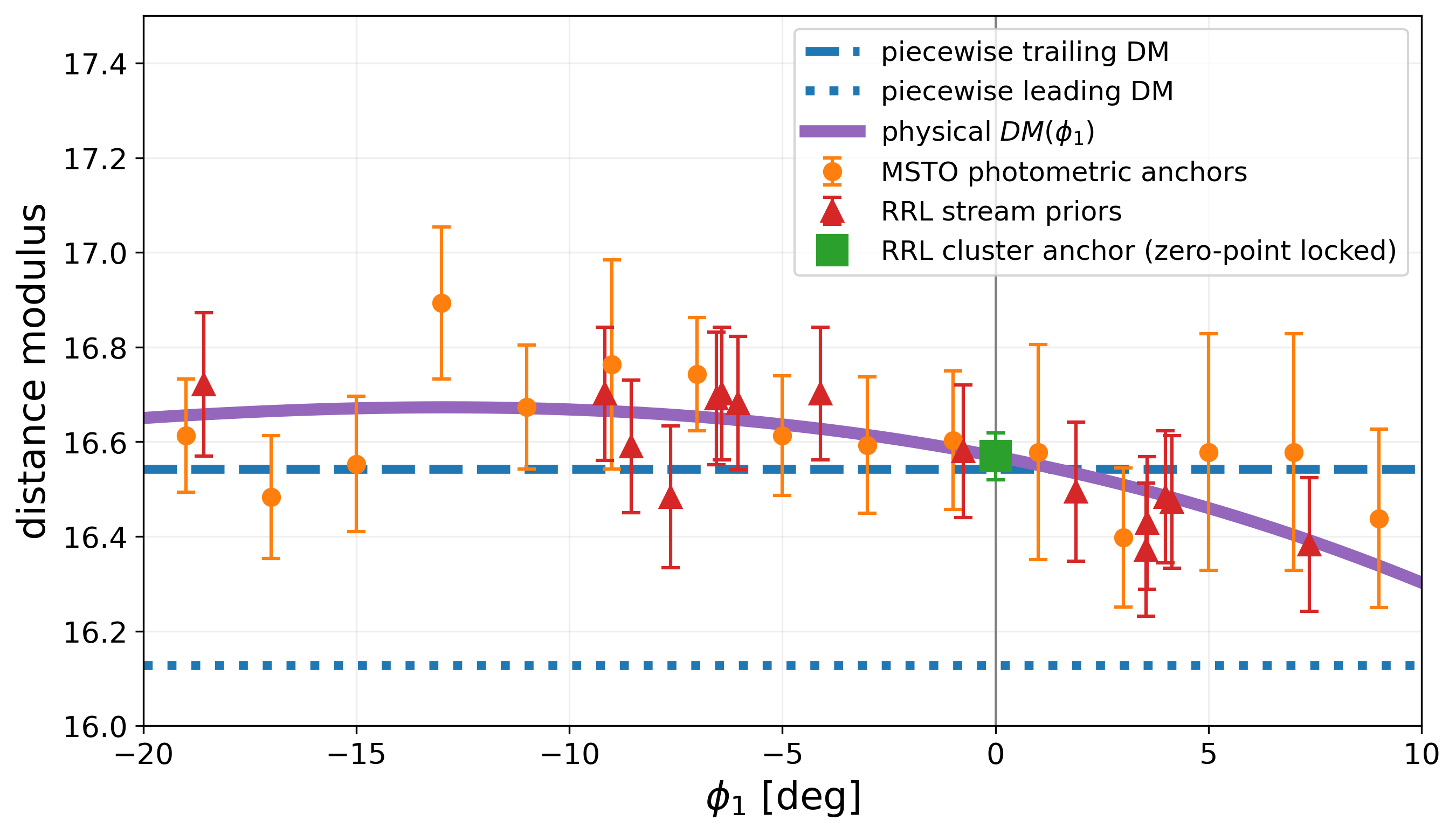}
\caption{DM curve used for the distance-dependent CMD rescreening. Orange circles show MSTO anchors from the background-subtracted CMD scans, calibrated to the RR Lyrae distance-modulus scale. Red triangles show stream RR Lyrae anchors, and the green square marks the cluster RR Lyrae zero-point anchor. Blue dashed and dotted horizontal lines indicate the first-pass trailing- and leading-arm CMD reference levels, converted to the same distance-modulus scale. The purple curve is the adopted physical $\text{DM}(\phione)$ relation fitted to the combined anchors; its relative variation shifts the aligned isochrone for the final CMD selection.}
\label{fig:dmtrack}
\end{figure*}

We retain the relative RR Lyrae distance offsets about ${\rm DM}_0$ and fit the calibrated MSTO and RR Lyrae anchors jointly with error weighting, normalizing the curve to ${\rm DM}(0)={\rm DM}_0$ (Figure~\ref{fig:dmtrack}). Hereafter, ${\rm DM}$ denotes the physical distance modulus, with the common $0.2\,{\rm kpc}$ RR Lyrae zero-point uncertainty. The adopted relation over $-20^\circ\le\phione\le10^\circ$ is

\begin{equation}
\begin{aligned}
\mathrm{DM}(x) &= 16.569 - 1.744\times10^{-2}x\\
&\quad - 8.394\times10^{-4}x^2\\
&\quad - 8.472\times10^{-6}x^3.
\end{aligned}
\label{eq:dmfunction}
\end{equation}

where $x=\phione/{\rm deg}$. For member selection, we shift the cluster-aligned isochrone by ${\rm DM}(\phione)-{\rm DM}_0$, using linear interpolation of the $0.1^\circ$ tabulation and holding the relation fixed beyond its boundaries. The resulting function is monotonic over most of the measured baseline: the leading arm is closer than the cluster, while the trailing arm is farther away in the adopted coordinate convention. The sense of this relative distance gradient agrees with the CFHT measurement of \citet{Ibata2016Pal5Photometry}.

\subsection{Depth-corrected stream morphology}
\label{subsec:morphmodel}

The continuous $\text{DM}(\phione)$ relation is then used to repeat the CMD selection over the full DR10 footprint. The resulting distance-rescreened sample provides the input for the final measurement of the projected \Pal\ morphology. We perform independent local fits in 41 overlapping $\phione$ windows centered at $\phi_{1,i}=-20^\circ,-19.25^\circ,\ldots,10^\circ$, separated by $0.75^\circ$. We adopt a full window width 1.5 times the center spacing. Each window therefore contains stars with $|\phione-\phi_{1,i}|\leq0.56^\circ$. The fitted quantities are the stream centroid $\mu_i$, the Gaussian cross-stream width $\sigma_i$, the stream amplitude, and the normalization and slope of the local field star background. The posterior stream component is then integrated over the local window to obtain the expected number of selected stream stars, $N_{{\rm stream},i}$.

The observed selected-star intensity contains three contributions: the response of the catalog to the local imaging depth, the spatially varying field star population, and the stream signal. To describe the spatial variation of the survey response, we divide the morphology footprint in the $(\phione,\phitwo)$ frame into $0.5^\circ\times0.5^\circ$ cells. For a position $\mathbf{x}=(\phione,\phitwo)$, we define $d_g(\mathbf{x})$ as the median $g$-band point-source depth in the cell containing $\mathbf{x}$. For the local $\phione$ window $i$, the observed and depth-normalized intensities are modeled as:
\begin{align}
\lambda_{{\rm obs},i}(\mathbf{x}) &=\eta_{\rm exp}[d_g(\mathbf{x})]\lambda_{{\rm int},i}(\phitwo), \nonumber\\
\lambda_{{\rm int},i}(\phitwo) &= A_{{\rm stream},i}\, \mathcal{N}_i(\phitwo\mid\mu_i,\sigma_i) + \nonumber\\
&\quad\ A_{{\rm bg},i}\, B_{{\rm lin},i}(\phitwo\mid b_i).
\label{eq:morphlambda}
\end{align}
Here $i$ labels the local $\phione$ window, $\lambda_{{\rm obs},i}$ is the expected observed selected-star intensity per unit solid angle, and $\lambda_{{\rm int},i}$ is the corresponding depth-normalized intrinsic intensity. The dimensionless function $\eta_{\rm exp}[d_g(\mathbf{x})]$ describes the response of the selected-star catalog to the local $g$-band limiting depth. The stream component is $A_{{\rm stream},i}\mathcal{N}_i(\phitwo\mid\mu_i,\sigma_i)$, where $A_{{\rm stream},i}$ sets the depth-normalized stream amplitude and $\mathcal{N}_i$ is a unit-normalized Gaussian cross-stream profile with centroid $\mu_i$ and dispersion $\sigma_i$. The field component is $A_{{\rm bg},i}B_{{\rm lin},i}(\phitwo\mid b_i)$, where $A_{{\rm bg},i}$ sets the local depth-normalized field-star intensity and $B_{{\rm lin},i}$ describes its smooth linear variation across $\phitwo$, with $b_i$ controlling the cross-stream slope.

We determine the exposure response from selected field stars in off-stream calibration cells. Among the $0.5^\circ\times0.5^\circ$ cells defined above, those lying outside the $1.5^\circ$ band around the first-pass stream track form the calibration sample. For each calibration cell $k$, we record the number of selected stars $N_k$, its unmasked angular area $\Omega_k$, and its median $\texttt{PSFDEPTH\_G}$ value $d_{g,k}$.

The calibration cells are grouped by $d_g$. In depth interval $\ell$, the selected field star surface density is
\begin{equation}
\widehat{\Sigma}_{{\rm field},\ell} = \frac{\displaystyle\sum_{k\in\ell}N_k}{\displaystyle\sum_{k\in\ell}\Omega_k}. \label{eq:bgresponse}
\end{equation}
A smooth one-dimensional function, $f_{\rm exp}(d_g)$, is fitted to the measured $\widehat{\Sigma}_{\rm field}$--$d_g$ relation. The normalized response is
\begin{equation}
\eta_{\rm exp}(d_g) = \frac{f_{\rm exp}(d_g)}{\left\langle f_{\rm exp}(d_g)\right\rangle_{\rm cal}}, \label{eq:eta_bg}
\end{equation}
where the denominator is the area-weighted mean over the calibration cells. The response is evaluated in every spatial cell of the morphology footprint. A region with $\eta_{\rm exp}>1$ is expected to contain more selected stars because its limiting depth is greater than the footprint average, while a region with $\eta_{\rm exp}<1$ has a reduced selected star yield.

\begin{figure*}[t]
\centering
\includegraphics[width=0.78\textwidth]{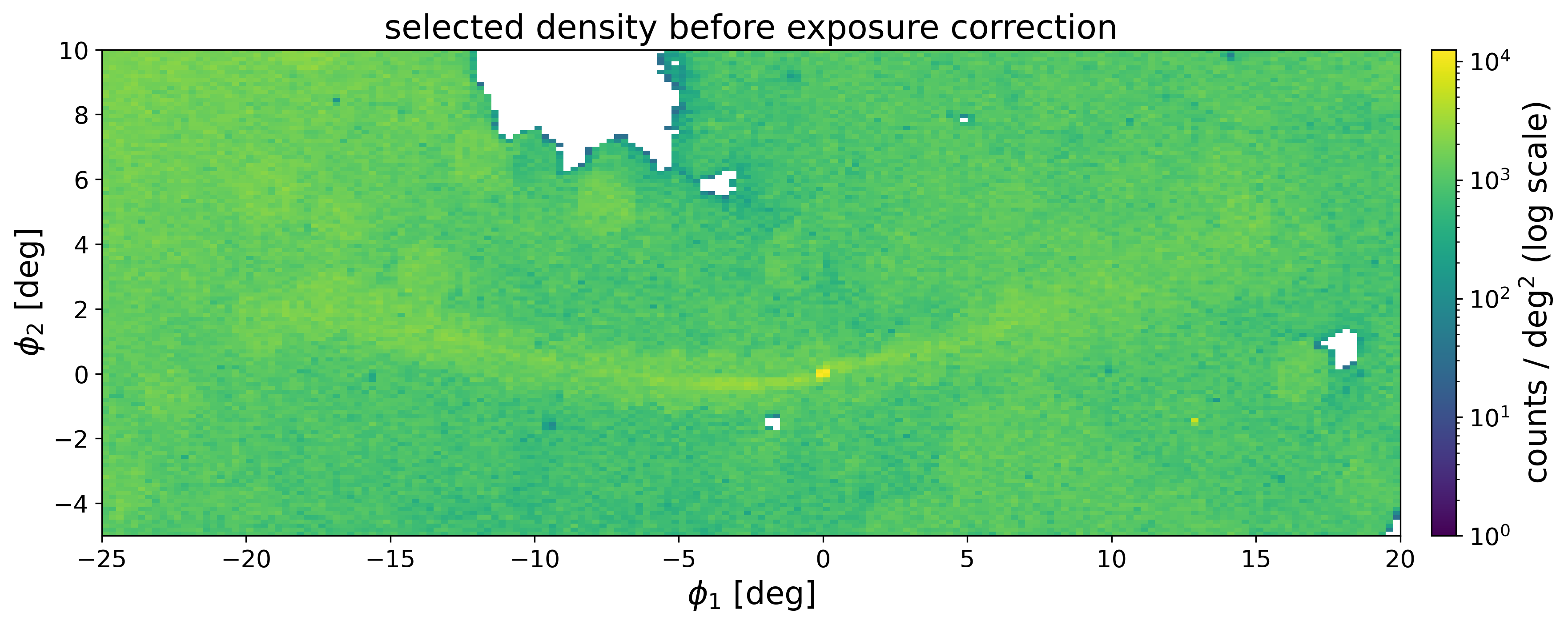}\\[2pt]
\includegraphics[width=0.78\textwidth]{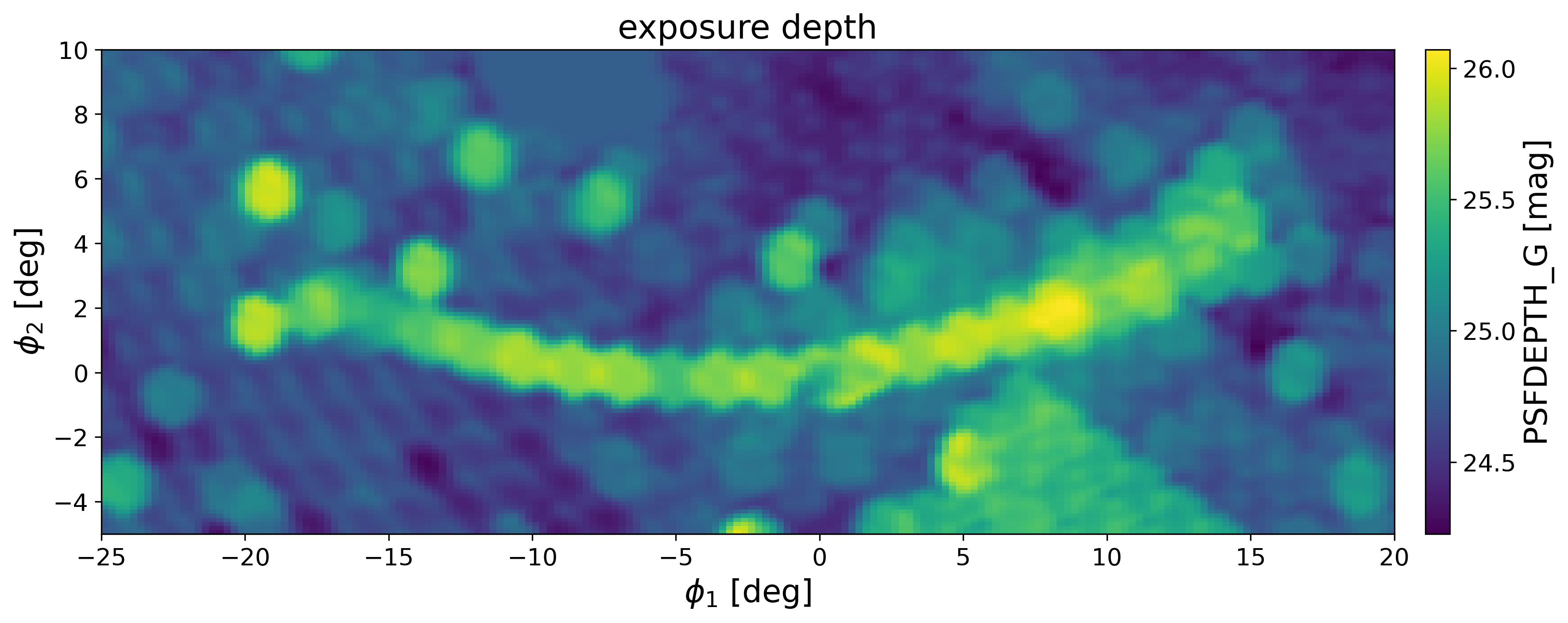}\\[2pt]
\includegraphics[width=0.78\textwidth]{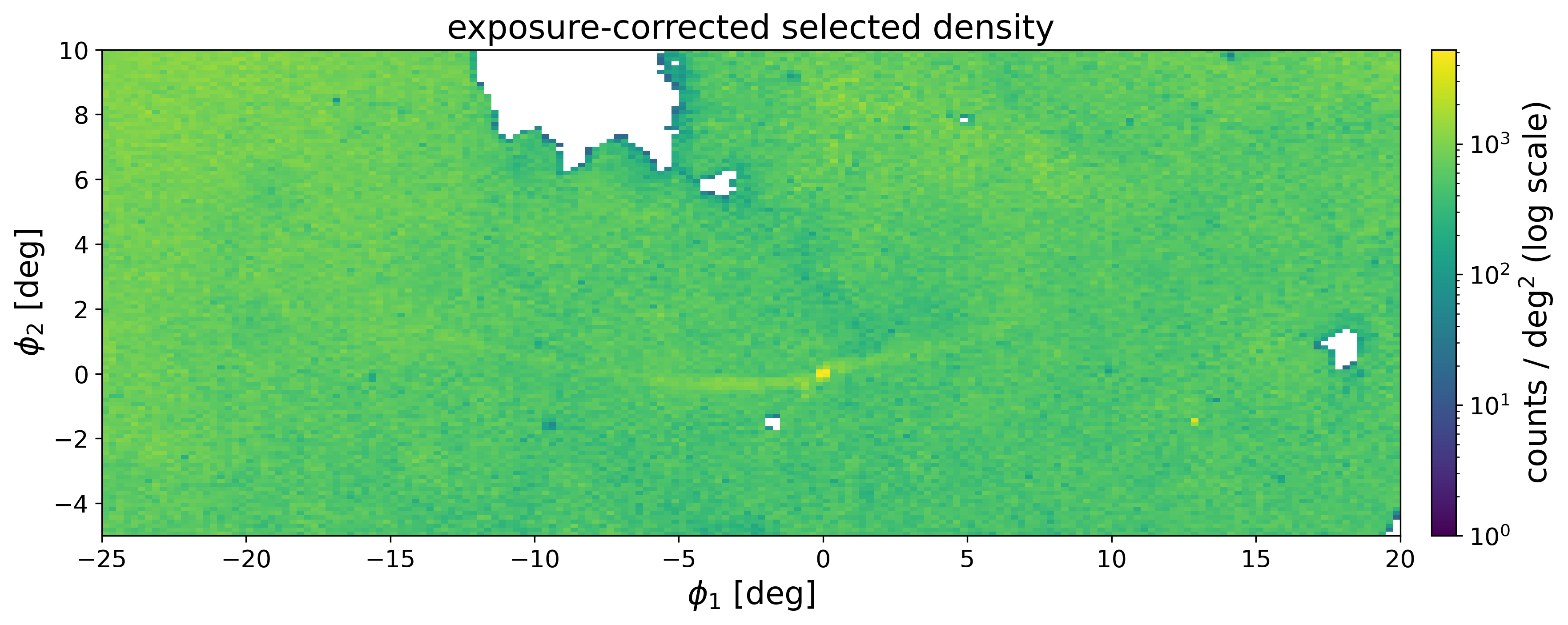}
\caption{Wide-field selected star density and depth response maps. Top: surface density of the distance-rescreened photometric sample before the depth correction. Middle: median $g$-band point-source depth, $d_g=\mathrm{median}(\texttt{PSFDEPTH\_G})$, in each spatial cell. Bottom: selected star density weighted by $\eta_{\rm exp}^{-1}(d_g)$.}
\label{fig:exposure}
\end{figure*}

Figure~\ref{fig:exposure} shows the depth calibration over the full search region. The uncorrected density map in the top panel contains broad variations that follow the spatial structure of the depth map in the middle panel. Weighting each selected star by $\eta_{\rm exp}^{-1}$ reduces these survey depth variations and provides the depth-normalized map in the bottom panel. The coherent stream now appears as the dominant feature following the known Palomar 5 locus, while the surrounding search region is dominated by low-level fluctuations.

After the exposure response has been fixed, the intrinsic field-star background is determined independently in each $\phione$ window. Its normalization $A_{{\rm bg},i}$ and cross-stream slope $b_i$ vary between windows, allowing both the foreground level and its local gradient to change along the stream.

The parameters in Equation~\ref{eq:morphlambda} are inferred with an extended Poisson likelihood. For the selected stars in $\phione$ window $i$ ($\mathcal{W}_i$),
\begin{equation}
\ln\mathcal{L}_i = \sum_{n\in\mathcal{W}_i} \ln\lambda_{{\rm obs},i} \!\left(\mathbf{x}_n\right) - \Lambda_i,
\end{equation}
where the expected total count is
\begin{equation}
\Lambda_i = \int_{\mathcal{W}_i} \lambda_{{\rm obs},i} \!\left(\mathbf{x}\right) \,d\Omega. \label{eq:morphnorm}
\end{equation}
The expected observed stream count is obtained by integrating the
exposure-modulated stream component over the same local fitting region,
\begin{equation}
N_{{\rm stream},i}=\int_{\mathcal{W}_i}\eta_{\rm exp}\!\left[d_g(\mathbf{x})\right]\, A_{{\rm stream},i}\, \mathcal{N}_i(\phitwo\mid\mu_i,\sigma_i)\, d\Omega.
\label{eq:nstream}
\end{equation}

We sample each local posterior with \texttt{emcee} \citep{ForemanMackey2013}. The centroid and width priors are centered on the smoothed first-pass morphology, $\mu_{0,i}$ and $\sigma_{0,i}$:
\begin{equation}
\ln p_{{\rm morph},i} = -\frac{1}{2} \left[\frac{(\mu_i-\mu_{0,i})^2}{(0.12^\circ)^2} + \frac{(\ln\sigma_i-\ln\sigma_{0,i})^2}{0.15^2}\right], \label{eq:morphprior}
\end{equation}
with $|\mu_i-\mu_{0,i}|<0.35^\circ$. The prior is applied independently in each $\phione$ window. The reported centroid, width, and stream count are the posterior medians, and their uncertainties are obtained from the 16th and 84th percentiles.

\begin{figure*}[t]
\centering
\includegraphics[width=0.88\textwidth]{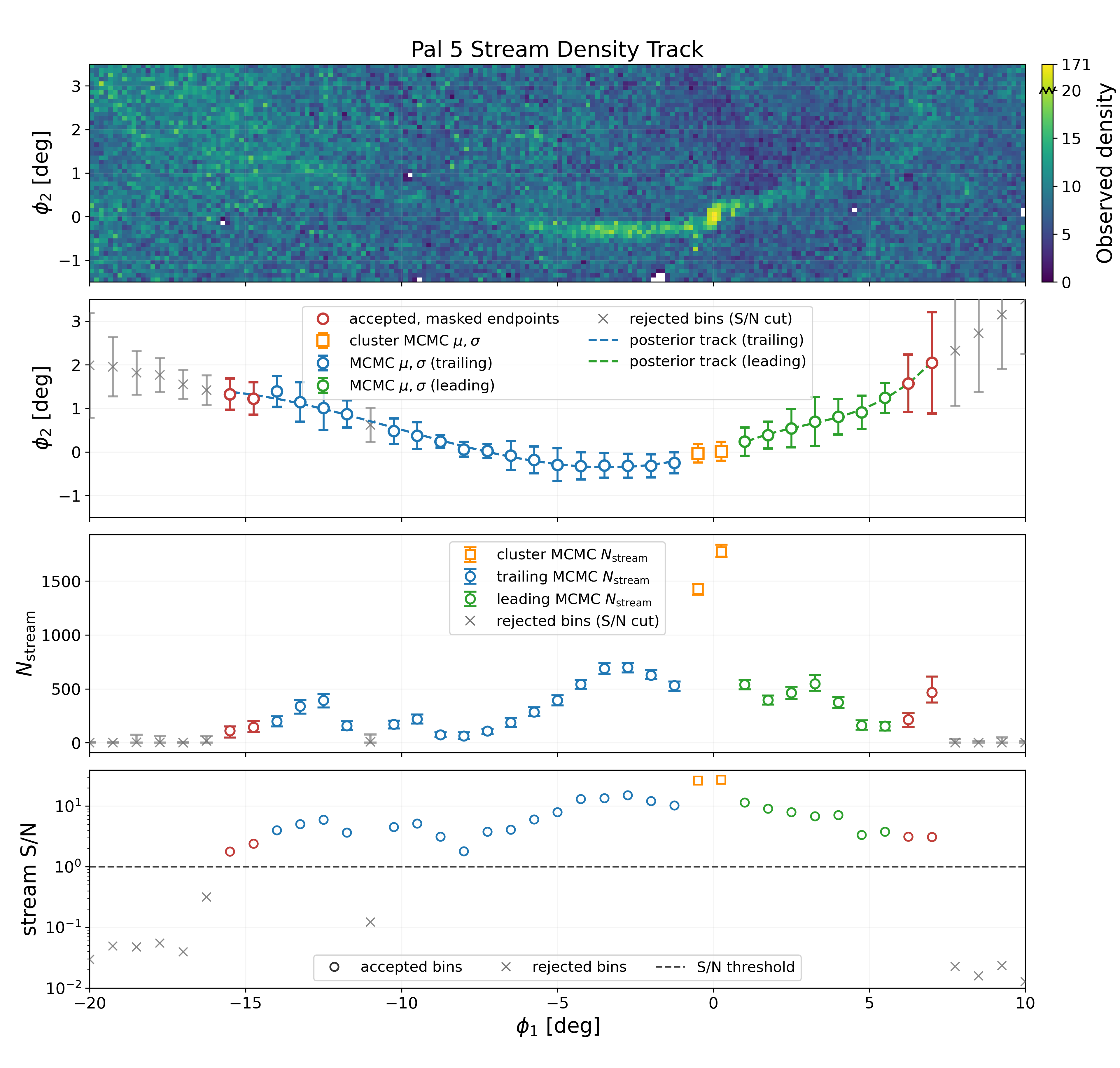}
\caption{Local morphology measurements from the distance-rescreened DR10 photometric sample. From top to bottom, the panels show the selected star density in stream coordinates, the fitted centroid sequence and width, the posterior stream counts, and the stream S/N. The circles and cluster-region squares mark $\phione$ windows satisfying ${\rm S/N}_{\rm stream}>1$, and the crosses mark lower-significance windows. Red circles identify the four tail-end bins excluded from the dynamical fit.}
\label{fig:densitytrack}
\end{figure*}

We quantify the significance of each local stream measurement with
\begin{equation}
{\rm S/N}_{{\rm stream},i} = \frac{N_{{\rm stream},i}}{\max\!\left(\delta N_{{\rm stream},i,-}, \delta N_{{\rm stream},i,+}\right)}, \label{eq:streamsnr}
\end{equation}
where $\delta N_{{\rm stream},i,-}$ and $\delta N_{{\rm stream},i,+}$ are the lower and upper posterior uncertainties on the stream count. Thirty of the 41 $\phione$ windows satisfy ${\rm S/N}_{\rm stream}>1$. Their centers span $-15.5^\circ\leq\phione\leq7.0^\circ$, with several lower-significance windows inside and outside this range retained in the full morphology table. The four tail-end bins identified by the perturbation mask remain part of the observed morphology measurement and appear in Figure~\ref{fig:densitytrack}. The fiducial potential likelihood in Section~\ref{sec:model} uses the remaining 26 bins.

\subsection{Extent of the significant DR10 morphology}
\label{subsec:null}

The response-corrected morphology is significant from $\phione=-15.5^\circ$ on the trailing side to $\phione=7.0^\circ$ on the leading side. Within this range the morphology follows a coherent centroid sequence, while the density varies substantially along the stream, including the depleted window near $\phione\simeq-11^\circ$. The stream bends from positive $\phitwo$ on the far trailing side toward the cluster and rises again toward positive $\phitwo$ on the leading side. The leading arm is broader and lower contrast than the inner trailing arm, so the detected leading extent is shorter under the same stream count criterion. This asymmetry agrees qualitatively with previous photometric and spectroscopic evidence for leading arm fanning.

The expanded footprint improves the empirical control of the contaminating field. The uncorrected density map contains large-scale gradients associated with depth and foreground structure; after applying the field star response, the known \Pal\ stream remains the only coherent high-contrast feature in the searched region. No additional outer tail is detected. This detection boundary is not a calibrated surface-brightness upper limit, and it is consistent with the caution from earlier \Pal\ studies that apparent clumps and gaps at low surface brightness are sensitive to completeness and foreground variations \citep{Thomas2016,Bonaca2020}.

\section{Mock Streams and Potential Inference}
\label{sec:model}

\subsection{Potential family and flattening convention}
\label{subsec:potential}

We interpret the observed morphology with mock streams generated using the \texttt{gala} package \citep{PriceWhelan2017Gala}. The present likelihood uses the projected stream centroid and cross-stream
width, which primarily probe the force field over the limited radial range spanned by \Pal\ and its visible debris. In this regime, the halo scale radius, normalization, and flattening are strongly degenerate: changing $r_s$ can be partly compensated by adjusting the enclosed mass while yielding a similar projected track. 

We therefore adopt a deliberately restricted baseline potential. The bulge and disk parameters are held fixed, the NFW scale radius is fixed at $r_s=20\,{\rm kpc}$, and the halo normalization is determined by the adopted circular-speed constraint at the Sun. Within the halo model, $\qphiz$ is the only free structural parameter. The Milky Way potential is composed of three components:
\begin{equation}
\Phi_{\rm MW}=\Phi_{\rm bulge}+\Phi_{\rm disk}+\Phi_{\rm halo}.
\end{equation}
The bulge is modeled as a Hernquist potential \citep{Hernquist1990},
\begin{equation}
\Phi_{\rm bulge}(r)=-\frac{G M_{\rm b}}{r+c_{\rm b}},
\end{equation}
with $M_{\rm b}=3.4\times10^{10}\,\msun$ and scale radius $c_{\rm b}=0.7\,{\rm kpc}$. The disk is a Miyamoto--Nagai potential \citep{MiyamotoNagai1975},
\begin{equation}
\Phi_{\rm disk}(R,z)=-\frac{G M_{\rm d}} {\left[R^2+\left(a_{\rm d}+\sqrt{z^2+b_{\rm d}^2}\right)^2\right]^{1/2}},
\end{equation}
with $M_{\rm d}=10^{11}\,\msun$, $a_{\rm d}=6.5\,{\rm kpc}$, and $b_{\rm d}=0.26\,{\rm kpc}$. Here $R=(x^2+y^2)^{1/2}$ is the Galactocentric cylindrical radius and $r=(R^2+z^2)^{1/2}$ is the spherical Galactocentric distance. The halo is an NFW potential \citep{Navarro1996}. The free parameter of the halo is the potential flattening. The spherical NFW potential is
\begin{equation}
\Phi_{\rm halo}=-\frac{G M_{\rm h}}{r_s}\frac{\ln(1+u)}{u},
\end{equation}
where the dimensionless radius is generalized to a triaxial form:
\begin{equation}
u=\frac{1}{r_s}\left[\left(\frac{x}{a}\right)^2+ \left(\frac{y}{b}\right)^2+\left(\frac{z}{c}\right)^2\right]^{1/2}. \label{eq:galaq}
\end{equation}
We set $a=1$, $b=1$, and $c=\qphiz$, placing the flattening inside the potential expression. Throughout this paper, $\qphiz$ denotes this halo potential flattening parameter. The mapping between $\qphiz$ and a density axis ratio depends on position and on the full potential via Poisson's equation. Therefore, comparisons to density flattening measurements require forward modeling under the same convention.

The halo mass normalization is fixed by imposing a total circular speed of $V_c(R_0)=232\,\kms$ \citep{McMillan2017} at $R_0=8.20\,{\rm kpc}$, adopting the Solar Galactocentric distance recommended by \citet{bland2016}. At each trial point in our parameter space, we solve for the NFW normalization such that the combined bulge+disk+halo radial acceleration satisfies
\begin{equation}
V_c^2(R_0)=-R_0\,a_R(R_0,z=0),
\end{equation}
where $a_R$ is the Galactocentric cylindrical radial component of the total acceleration,
\begin{equation}
a_R=\frac{x a_x+y a_y}{R},
\end{equation}
evaluated in the midplane at $(R_0,z=0)$ for the circular speed constraint. The circular speed constraint anchors the mass normalization of the inner potential while allowing the halo flattening and the \Pal\ phase-space coordinates to determine the projected stream morphology. The derived NFW mass is recorded as a diagnostic of this fixed-$V_c$ model.

\subsection{Mock stream reduction and likelihood}
\label{subsec:likelihood}

For each trial potential and present-day \Pal\ phase-space point, we generate a mock stream using the modified Fardal-style particle release machinery implemented in \texttt{gala} \citep{Fardal2015,PriceWhelan2017Gala}, following the Lagrange-point stream-modeling approach pioneered by \citet{Varghese2011Streams}. The fiducial progenitor is represented by a Plummer model with mass $M_{\rm prog}=2\times10^4\,\msun$ and scale radius $b_{\rm prog}=15\,{\rm pc}$, adopting the mass and scale used by \citet{Erkal2017}. These parameters set the scale of the tracer stream used for the morphology comparison and are held fixed throughout the fit. Section~\ref{subsec:progenitorsensitivity} examines the dependence of the inferred halo shape on the progenitor mass and scale radius.

After generation, each mock stream is transformed to the \Pal\ coordinate frame and processed using the same $\phione$ window adopted for the observed morphology in Section~\ref{subsec:morphmodel}. At each window $\phi_{1,i}$, we select mock particles within $|\phione-\phi_{1,i}|\le0.56^\circ$ and compute the median $\phitwo$ as $\mu_{{\rm mock},i}$. The robust width is $\sigma_{{\rm mock},i}=0.7413\,\mathrm{IQR}(\phitwo)$, where the factor $0.7413$ converts the interquartile range to the Gaussian-equivalent standard deviation.

The sampled parameters are $\theta=(\qphiz,D,\mu_{\alpha *},\mu_\delta,v_{\rm los})$. We adopt a uniform prior on the halo potential flattening over $0.6<\qphiz<1.4$. The phase-space priors are Gaussian:
\begin{align}
D &= 20.6\pm0.2\,{\rm kpc},\\
\mu_{\alpha *} &= -2.754\pm0.017\,\masyr,\\
\mu_{\delta} &= -2.643\pm0.014\,\masyr,\\
v_{\rm los} &= -58.61\pm0.15\,\kms,
\end{align}
with a correlation coefficient of 0.01 between the two proper motion components. The distance prior follows the RR Lyrae scale of \citet{PriceWhelan2019}, and the proper motion and radial velocity priors are tied to recent globular cluster phase-space compilations based on Gaia EDR3 \citep{VasilievBaumgardt2021}.

The fiducial likelihood uses the S/N-accepted morphology measurements within $-14.5^\circ<\phi_{1,i}<6.0^\circ$, corresponding to 26 local morphology bins. Four additional measurements satisfy the stream-count significance criterion but lie outside this fiducial angular range, at $\phi_{1,i}=-15.5^\circ,\,-14.75^\circ,\,6.25^\circ,\,{\rm and}\,7.0^\circ$. These low-contrast tail-end bins show coherent departures from the main track and coincide with angular ranges where bar torques, fanning, and external perturbations have been emphasized in previous \Pal\ work \citep{Pearson2017,Erkal2017,Bonaca2020,Kuzma2022}.

The stream width is sensitive to the assumed progenitor structure, stripping history, background calibration, and unresolved perturbations. We therefore assign the width information a relative weight $w_\sigma=0.25$. The total morphology likelihood is defined as
\begin{equation}
\ln\mathcal{L}_{\rm morph}=\ln\mathcal{L}_{\mu}+w_\sigma\ln\mathcal{L}_{\sigma,0},
\label{eq:totallikelihood}
\end{equation}
where $\mathcal{L}_{\mu}$ is the centroid likelihood and
$\mathcal{L}_{\sigma,0}$ is the unweighted width likelihood.
For the centroid term, we define
\begin{equation}
V_{\mu,i}=s_{\mu,i}^2+s_{\mu,{\rm jit}}^2,
\end{equation}
and evaluate
\begin{equation}
\ln\mathcal{L}_{\mu}=-\frac{1}{2}\sum_i\left[\frac{(\mu_i-\mu_{{\rm mock},i})^2}{V_{\mu,i}}+\ln\left(2\pi V_{\mu,i}\right)\right].
\label{eq:mulikelihood}
\end{equation}
For the width term, we similarly define
\begin{equation}
V_{\sigma,i}=s_{\sigma,i}^2+s_{\sigma,{\rm jit}}^2,
\end{equation}
and
\begin{equation}
\ln\mathcal{L}_{\sigma,0}=-\frac{1}{2}\sum_i\left[\frac{(\sigma_i-\sigma_{{\rm mock},i})^2}{V_{\sigma,i}}+\ln\left(2\pi V_{\sigma,i}\right)\right].
\label{eq:sigmalikelihood}
\end{equation}
Here $s_{\mu,i}$ and $s_{\sigma,i}$ are the mean observational uncertainties from the local morphology posteriors.

We adopt fixed model-discrepancy floors of $s_{\mu,{\rm jit}}=0.05^\circ$, $s_{\sigma,{\rm jit}}=0.08^\circ$, chosen to be comparable to the residual centroid and width scatter found in preliminary smooth-potential fits. These terms prevent bins with very small formal posterior uncertainties from dominating the likelihood and account approximately for residual model mismatch. 

The sampled posterior is
\begin{equation}
\ln p(\boldsymbol{\theta}\mid\mathcal{D}_{\rm fid})=\ln\mathcal{L}_{\rm morph}+\ln p(\boldsymbol{\theta})+{\rm const.},
\end{equation}
where $p(\boldsymbol{\theta})$ contains the halo-shape and \Pal\ phase-space priors described above. The progenitor mass and scale radius are held constant throughout the 3 Gyr modeled stripping interval. $M_{\rm prog}$ represents an effective mass for particle release. We sample this posterior with \texttt{emcee}, using 64 walkers, 256 burn-in steps, and 1250 production steps per walker.

\subsection{Posterior constraint on the potential flattening}
\label{subsec:qresult}

Table~\ref{tab:mcmc} summarizes the marginalized posterior for the fiducial fit. The halo potential flattening parameter is

\begin{equation}
\qphiz=0.824^{+0.133}_{-0.110}, \label{eq:qresult}
\end{equation}

where the uncertainties are the 16th and 84th percentiles. The corresponding total potential force diagnostic at the \Pal\ position is

\begin{equation}
q_{\Phi,{\rm total}}({\rm Pal~5})=0.862^{+0.064}_{-0.053}.
\end{equation}

Here $\qphiz$ is the input halo potential flattening in Equation~\ref{eq:galaq}, whereas $q_{\Phi,{\rm total}}=[(z/R)(a_R/a_z)]^{1/2}$ is a local force-ratio diagnostic. The quantities $R$ and $z$ are the Galactocentric cylindrical radius and vertical coordinate, and $a_R$ and $a_z$ are the corresponding components of the total bulge+disk+halo acceleration. All quantities are evaluated at the present-day \Pal\ position for each posterior sample. The two values differ because the disk and bulge contribute non-negligibly to the total acceleration at the Galactocentric distance of \Pal.

\begin{table}[t]
\centering
\caption{MCMC posterior constraints for the fiducial perturbation-masked mock track fit with $M_{\rm prog}=20,000\,\msun$ and $b_{\rm prog}=15\,{\rm pc}$. Values are medians with 16th--84th percentile uncertainties.}
\label{tab:mcmc}
\begin{tabular}{lcc}
\hline\hline
Parameter & Posterior & Unit \\
\hline
$\qphiz$ & $0.824^{+0.133}_{-0.110}$ & -- \\
$D$ & $20.784^{+0.094}_{-0.114}$ & kpc \\
$\mu_{\alpha *}$ & $-2.714^{+0.009}_{-0.009}$ & mas yr$^{-1}$ \\
$\mu_\delta$ & $-2.679^{+0.007}_{-0.007}$ & mas yr$^{-1}$ \\
$v_{\rm los}$ & $-58.606^{+0.105}_{-0.108}$ & km s$^{-1}$ \\
\hline
$q_{\Phi,{\rm total}}({\rm Pal~5})$ & $0.862^{+0.064}_{-0.053}$ & -- \\
\hline
\end{tabular}
\end{table}

Table~\ref{tab:fitvariants} compares the fiducial inference with three diagnostic variants. Including all four tail-end bins shifts the posterior median from $q_z=0.824$ to $q_z=0.774$, although the credible intervals overlap. This change demonstrates that the coherent endpoint residuals have appreciable leverage on the curvature that the static model attributes to the halo shape. By contrast, removing the width term gives $q_z=0.833^{+0.094}_{-0.107}$, and increasing the stream-count threshold to ${\rm S/N}_{\rm stream}>3$ gives $q_z=0.830^{+0.137}_{-0.119}$. Both values are close to the fiducial result. The inferred flattening is therefore relatively insensitive to the downweighted width information and to the inclusion of the marginal S/N morphology bins but is more sensitive to the treatment of the coherent tail-end residuals.

\begin{table*}[t]
\centering
\caption{Potential fit configurations for the fiducial analysis and diagnostic sensitivity tests. All fits use $M_{\rm prog}=20,000\,\msun$ and $b_{\rm prog}=15\,{\rm pc}$.}
\label{tab:fitvariants}
\begin{tabular}{llcc}
\hline\hline
Fit & Bin selection and observables & $N_{\rm bin}$ & $q_z$ \\
\hline
fiducial & ${\rm S/N}_{\rm stream}>1$, excluding four tail-end bins; $\mu,\sigma$ & 26 & $0.824^{+0.133}_{-0.110}$ \\
no perturbation mask & ${\rm S/N}_{\rm stream}>1$, no perturbation-mask exclusion; $\mu,\sigma$ & 30 & $0.774^{+0.148}_{-0.103}$ \\
centroid-only & same bins as fiducial; $\mu$ only & 26 & $0.833^{+0.094}_{-0.107}$ \\
high-S/N & ${\rm S/N}_{\rm stream}>3$, excluding four tail-end bins; $\mu,\sigma$ & 25 & $0.830^{+0.137}_{-0.119}$ \\
\hline
\end{tabular}
\end{table*}

\begin{figure*}[t]
\centering
\includegraphics[width=0.82\textwidth]{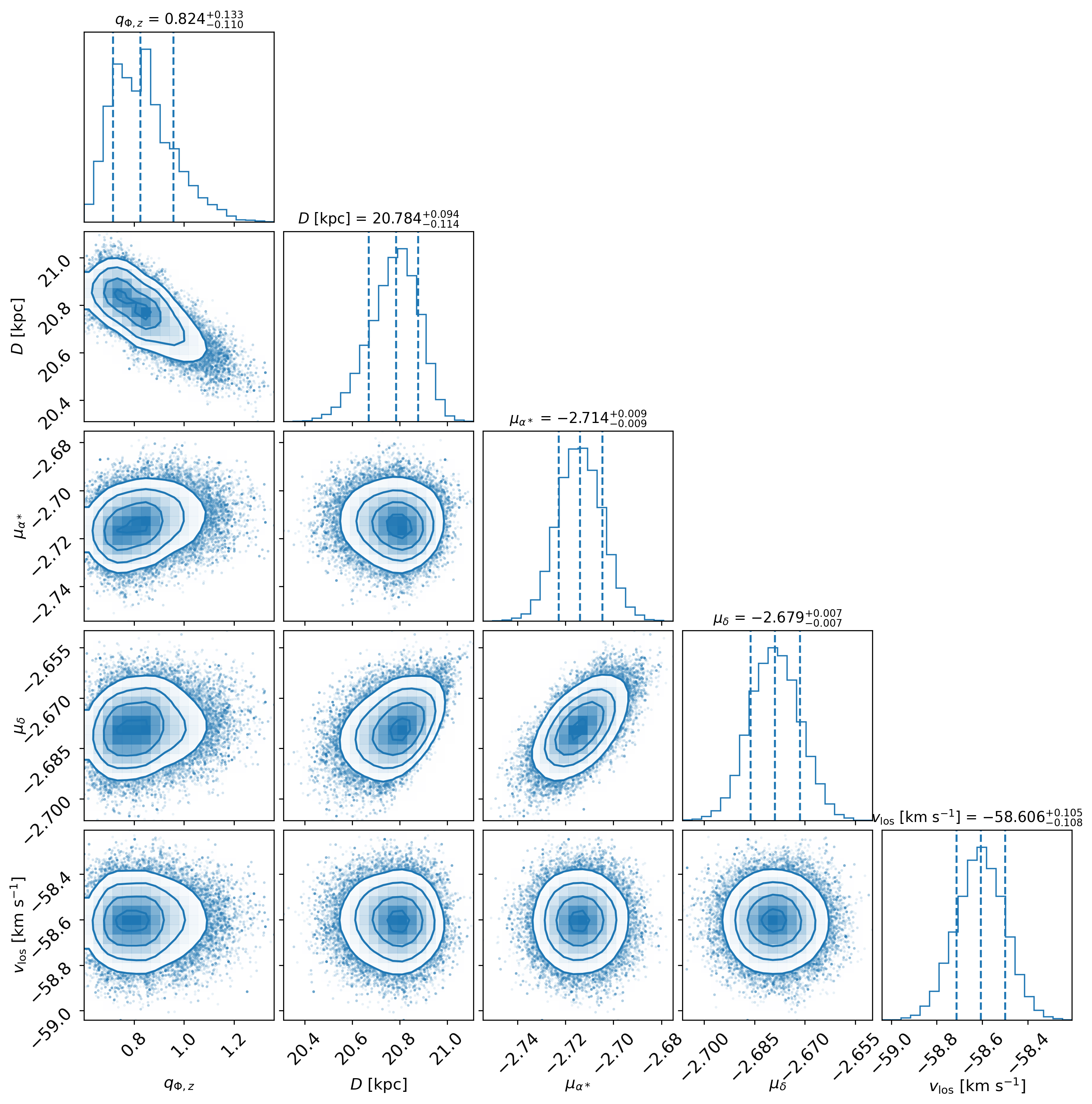}
\caption{Posterior distribution for the fiducial 26-bin perturbation-masked mock track fit. The sampled parameters are $\qphiz$, heliocentric distance, two proper motion components, and radial velocity. The halo normalization is solved at every step to enforce $V_c(R_0)=232\,\kms$, and $r_s$ is fixed at 20 kpc. The progenitor has $M_{\rm prog}=20,000\,\msun$ and $b_{\rm prog}=15\,{\rm pc}$. The inferred halo flattening is conditional on this potential family and on the adopted \Pal\ phase-space priors.}
\label{fig:corner}
\end{figure*}

Figure~\ref{fig:corner} shows the posterior distribution for the fiducial configuration. The strongest correlations involve $q_z$ and the present-day \Pal\ phase-space coordinates because changes in the latter also modify the projected stream curvature. The tight external phase-space priors keep the posterior compact and make the halo-shape result conditional on those priors. The posterior proper motions are displaced from the centers of the external priors by approximately $2$--$3$ prior standard deviations, indicating mild tension between the adopted smooth-potential family and the photometric track when the photometric track is fit without along-stream proper-motion or radial-velocity measurements in the likelihood.

\subsection{Track, width, and residuals}
\label{subsec:residuals}

\begin{figure*}[t]
\centering
\includegraphics[width=0.94\textwidth]{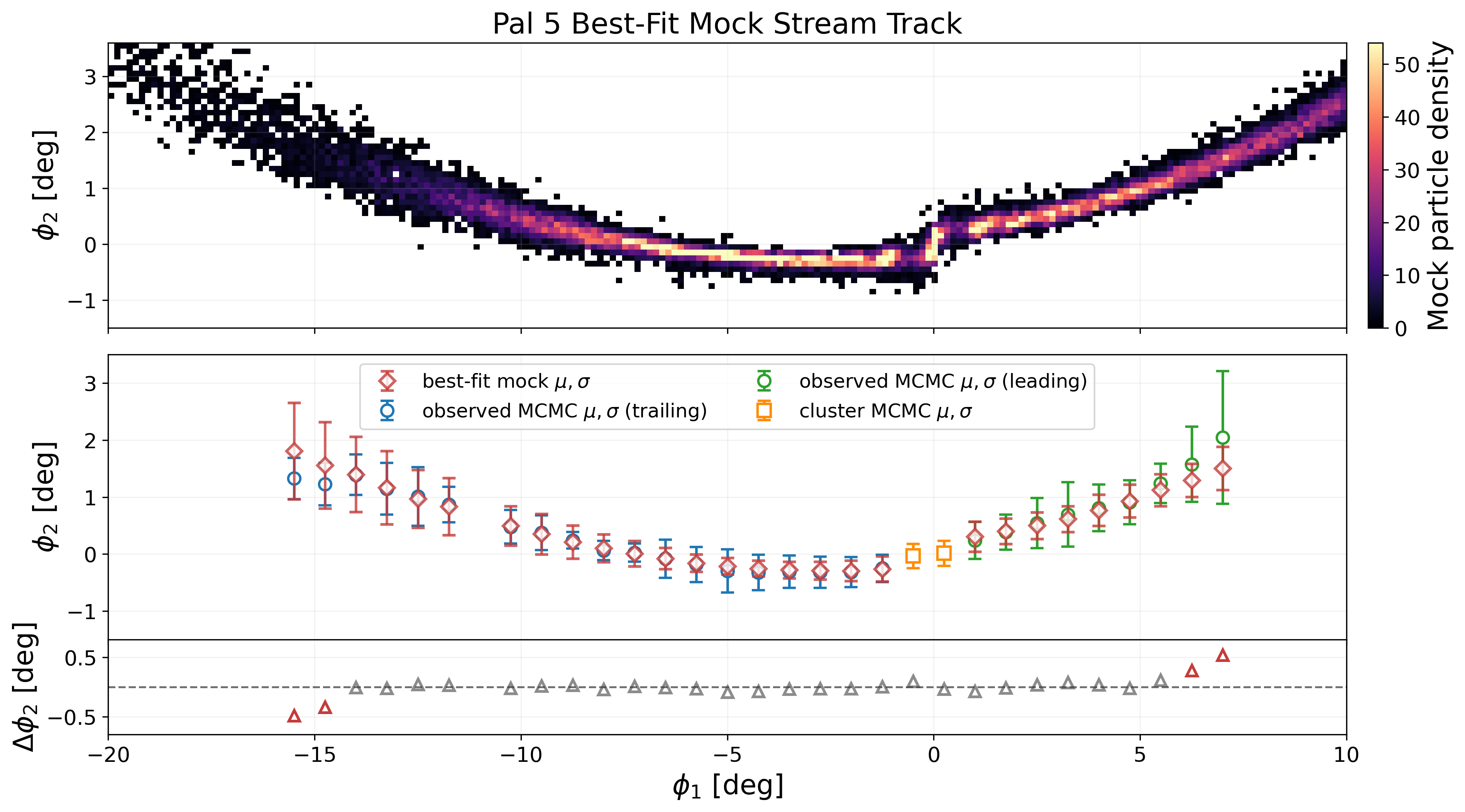}
\caption{Maximum-posterior mock stream compared with the observed morphology. The panels show the mock density in stream coordinates, the observed and mock centroid tracks with cross-stream widths, and the centroid residuals for the S/N-selected bins. Gray points are the bins included in the fiducial track-width likelihood. Red tail-end points are the four S/N-selected tail-end bins excluded by the perturbation mask; they are shown to illustrate coherent residuals relative to the static best-fit mock.}
\label{fig:bestfit}
\end{figure*}

Figure~\ref{fig:bestfit} compares the observed morphology with the maximum-posterior mock stream. The S/N-accepted centroid sequence traces a pronounced curved path. Along the trailing arm, the centroid decreases from $\mu\simeq1.33^\circ$ at $\phione=-15.5^\circ$
to a minimum of $\mu\simeq-0.32^\circ$ near $\phione\simeq-4^\circ$. It then returns toward $\phitwo=0$ near the cluster and rises along the leading arm, reaching $\mu\simeq2.05^\circ$ at $\phione=7^\circ$. The measured width in the high-contrast cluster region has a median $\sigma\simeq0.13^\circ$ for $|\phione|\lesssim2^\circ$, and increases toward the leading side, where the median width is $\simeq0.21^\circ$ and the outermost accepted bin reaches $\sigma\simeq0.58^\circ$. The stream count S/N follows the same qualitative pattern: the cluster bins have the highest contrast, whereas the outer leading and trailing bins are accepted at lower significance.

The maximum-posterior mock captures the main projected curvature after the four tail-end bins are excluded from the likelihood. For the 26 bins used in the halo fit, the centroid residuals have an RMS of $0.049^\circ$ and a maximum absolute value of $0.12^\circ$; the width residuals have an RMS of $0.080^\circ$. The four masked bins show much larger coherent departures. The trailing bins at $\phione=-15.5^\circ$ and $-14.75^\circ$ lie below the mock by $0.48^\circ$ and $0.33^\circ$, respectively, while the leading bins at $\phione=6.25^\circ$ and $7.0^\circ$ lie above the mock by $0.28^\circ$ and $0.54^\circ$. These offsets exceed the local centroid uncertainties and show a coherent sign reversal between the two tails, indicating end-of-tail departures rather than random scatter around the smooth model.

The four masked bins carry strong curvature leverage, so including them would shift the inferred effective flattening toward a model that absorbs localized tail-end structure into the halo shape. We therefore treat them as perturbed or selection-sensitive angular ranges and report them as residual diagnostics. The best-fit stream provides a static potential approximation to the unmasked main track; the excluded tail-end departures require a dynamical model with additional perturbations or a more complete selection function treatment.

\subsection{Sensitivity to progenitor mass and scale}
\label{subsec:progenitorsensitivity}

The progenitor mass during the formation of the observed tidal tails is uncertain. \citet{Ibata2017Pal5Mass} estimated a present-day bound-remnant mass of $4297\pm98\,\msun$ and a mass of $12,200\pm400\,\msun$ for the cluster plus visible stream, with initial-mass estimates ranging from approximately $12,200$ to $47,000\,\msun$ depending on the assumed original mass function. Motivated by these mass scales, we fit mock streams with fixed effective masses of $12,200$, $20,000$, and $47,000\,\msun$ at $b_{\rm prog}=15\,{\rm pc}$, and include a model with $M_{\rm prog}=20,000\,\msun$ and $b_{\rm prog}=4\,{\rm pc}$. All models use the same 26 morphology bins, Galactic potential, phase-space priors, and particle-release and likelihood settings.

\begin{figure*}[t]
\centering
\includegraphics[width=0.98\textwidth]{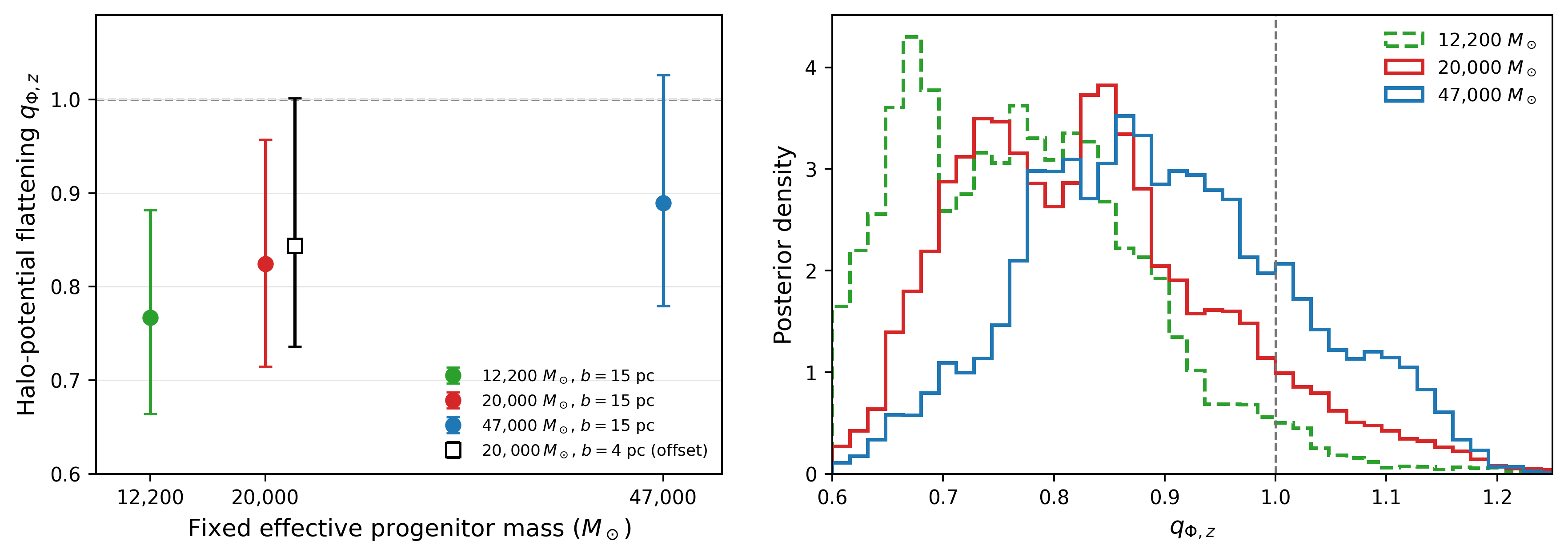}
\caption{Dependence of the inferred halo potential flattening on the progenitor mass and Plummer scale radius. Left: posterior medians and central 68\% intervals for three masses at $b_{\rm prog}=15\,{\rm pc}$. The black open square denotes the model with $M_{\rm prog}=20,000\,\msun$ and $b_{\rm prog}=4\,{\rm pc}$, offset horizontally for clarity. Right: marginalized halo-flattening posteriors for the three 15 pc models, with the same colors as in the left panel. Gray dashed lines mark a spherical halo potential. Each posterior is conditional on its fixed progenitor mass and scale radius.}
\label{fig:progenitorsensitivity}
\end{figure*}

Table~\ref{tab:progenitorsensitivity} and Figure~\ref{fig:progenitorsensitivity} show that the posterior median increases with the adopted mass, from $\qphiz=0.767$ through $0.824$ to $0.889$. Relative to the fiducial model, the lower- and higher-mass models shift the median by $-0.057$ and $+0.065$, respectively. At $M_{\rm prog}=20,000\,\msun$, increasing the Plummer scale from 4 to 15 pc changes the median by $-0.019$. The scale dependence is therefore smaller than the statistical intervals and the mass dependence over the adopted range. The intervals overlap and are conditional on each fixed progenitor model; the $47,000\,\msun$ model includes a spherical halo potential within its central 68\% interval.

The different sensitivities have a qualitative interpretation within the adopted particle-release model. At fixed orbit and Galactic tidal field, the tidal length scale varies as $M_{\rm prog}^{1/3}$, and the release-velocity scale varies with it \citep{Fardal2015,PriceWhelan2017Gala}. Changing the progenitor mass therefore changes the debris offsets and subsequent stream morphology, introducing a degeneracy with the halo shape and progenitor phase-space coordinates. The Plummer radius instead changes the progenitor force acting on the particles. Well outside this radius, the force depends mainly on the total mass, providing a possible explanation for the weaker scale-radius dependence over the range considered here.

\begin{table}[t]
\centering
\caption{Sensitivity of the halo potential flattening to the fixed progenitor mass and Plummer scale. All fits use the fiducial 26 bins. Values are posterior medians with 16th--84th percentile uncertainties; $\Delta q_{z,50}$ is the median difference from the fiducial $20,000\,\msun$, 15 pc model.}
\label{tab:progenitorsensitivity}
\begin{tabular}{rrcr}
\hline\hline
$M_{\rm prog}$ ($\msun$) & $b_{\rm prog}$ (pc) & $\qphiz$ & $\Delta q_{z,50}$ \\
\hline
12,200 & 15 & $0.767^{+0.115}_{-0.103}$ & $-0.057$ \\
20,000 & 15 & $0.824^{+0.133}_{-0.110}$ & $0.000$ \\
47,000 & 15 & $0.889^{+0.137}_{-0.110}$ & $+0.065$ \\
20,000 & 4 & $0.844^{+0.158}_{-0.108}$ & $+0.019$ \\
\hline
\end{tabular}
\end{table}

\section{Discussion}
\label{sec:discussion}

\subsection{Wide-field morphology and outer tail sensitivity}
\label{subsec:discmorph}

The significant DR10 morphology measured here overlaps the high-confidence stream traced in earlier SDSS, CFHT, and DECaLS analyses \citep{Odenkirchen2003,Ibata2016Pal5Photometry,Erkal2017,Bonaca2020}. The trailing side is detected over a longer angular baseline than the leading side, as in the original SDSS maps \citep{Odenkirchen2003,GrillmairDionatos2006,zou2009tidal}. The leading arm remains broader and more diffuse, consistent with the fanning emphasized by \citet{Bonaca2020}, supported kinematically by \citet{Kuzma2022}, and consistent with the low expected yield of bright red giant members in diffuse tail pointings \citep{Sheffield2025}. The present analysis extends the search footprint and applies a single distance-rescreened CMD selection and exposure response across the field. Within this selection, the response-corrected stream is significant over $-15.5^\circ\le\phione\le7.0^\circ$, and no additional coherent outer continuation is detected.

The CFHT studies provide deep, high-quality photometric constraints on the stream track and relative distance gradient, together with measurements of its density and width \citep{Ibata2016Pal5Photometry,Ibata2017Pal5Mass,Erkal2017}. The complementary advantage of DR10 is its extensive off-stream coverage. We use the surrounding field to calibrate the selected-star response to imaging depth, while fitting the local background normalization and gradient in each stream window (Section~\ref{subsec:morphmodel}). This wider spatial context helps distinguish the low-contrast outskirts from foreground and depth variations, complementing the detailed CFHT measurements.

The non-detection should be interpreted as a detection boundary under the present photometric selection, depth response, and S/N threshold. The outer tails are measured against a structured Milky Way foreground, and the apparent density field depends on photometric depth, the color--color filter, the distance-dependent CMD mask, and the local background model. The uncorrected and corrected maps in Figure~\ref{fig:exposure} show that depth variations are large enough to affect visual interpretation of the outskirts. A quantitative upper limit on undetected debris requires artificial stream recovery: injected streams should span surface brightness, width, distance gradient, and stellar population and should be passed through the same $g,r,z$ cuts, $\text{DM}(\phione)$-dependent CMD filter, exposure correction, local likelihood, and S/N threshold.

\subsection{Coherent residuals at the stream ends}
\label{subsec:discresiduals}

The residual panel of Figure~\ref{fig:bestfit} shows that the masked end points form a coherent pattern rather than uncorrelated scatter around the maximum-posterior static model. The far trailing bins lie below the mock track, while the far leading bins lie above it. Such a sign reversal would give the masked bins strong curvature leverage on $q_z$ if they were included in the likelihood; these same bins occupy the lowest-contrast and most perturbation-sensitive parts of the measured morphology.

Similar track-level residuals have motivated several physical interpretations in previous work. \citet{Bonaca2020} measured variations in the width, density, and direction of the \Pal\ tails and found that a rotating bar can reproduce some of the density and width structure, while the trailing arm track remained difficult to explain in their fiducial model. Bar-driven asymmetries and direction changes are also expected in simulations of \Pal-like streams \citep{Pearson2017}. A tilting Galactic disk can generate fanning and width variations that mimic halo shape signatures in static models \citep{Nibauer2024}. Encounters with dark subhalos or known globular clusters can perturb selected sections of the debris \citep{Erkal2017,Ferrone2025}. Analogous behavior appears in other thin streams: the GD-1 spur and gap can be produced by an impulsive encounter that places debris on an offset loop \citep{Bonaca2019GD1}, and the GD-1 wiggle near density gaps provides a second example of coherent track residuals \citep{deBoer2020}. The \Pal\ tail-end residuals therefore provide diagnostics of perturbations and disequilibrium in addition to the smooth halo shape.

\subsection{Halo shape measurement in context}
\label{subsec:discshape}

The mock stream posterior gives $q_z=0.824^{+0.133}_{-0.110}$ for the halo potential flattening parameter in the adopted potential-flattened NFW convention. The posterior median is more oblate than several \Pal-focused estimates, although the intervals are broad. \citet{Kupper2015} modeled \Pal\ with streakline streams and obtained $q_z=0.95^{+0.16}_{-0.12}$ for the halo potential. \citet{Bovy2016} used \Pal\ and GD-1 phase-space tracks to infer total potential flattenings of $0.94\pm0.05$ at \Pal\ and $0.95\pm0.04$ at GD-1, together with an inner 20 kpc dark matter density axis ratio of $1.05\pm0.14$. \citet{Palau2023} found $q_{\rho}^{\rm h}=1.01\pm0.09$ from \Pal\ alone and $q_{\rho}^{\rm h}=1.06\pm0.06$ from a joint fit to \Pal, M68, and NGC~3201. Taken at face value, the published point estimates are separated. Their statistical compatibility cannot be assessed directly, however, because the analyses constrain different physical quantities and do not use a common data set or forward model.

The main differences fall into several categories. The data sets range from SDSS-era star counts and early \Pal\ tail tracks to CFHT, DECaLS, DR10, Gaia astrometry, RR Lyrae distances, and spectroscopic velocities. The fitted observables also differ: some analyses fit only projected tracks, others include distance, proper motion, radial velocity, density, or multiple streams, while the fiducial fit here uses the DR10 centroid and width for the stream morphology. The potential families range from static axisymmetric halos to triaxial or acceleration field descriptions, and most published halo shape measurements differ in whether the Galactic bar, the LMC, disk disequilibrium, subhalos, or globular cluster encounters are included. The flattening parameters are likewise not interchangeable: the $q_z$ in this work and in some \Pal\ streakline models is a potential shape parameter, \citet{Bovy2016} quoted total potential flattening and a derived density axis ratio, \citet{Palau2023} reported halo density flattening, and direct acceleration analyses constrain the shape of the acceleration field. A comparison at the level of ``oblate'' or ``near spherical'' is informative, but a one-to-one numerical comparison requires rerunning the different data sets in a common forward model.

Within that context, our result should be read as an effective constraint on the static smooth potential required to fit the main \Pal\ track and width. The perturbation mask removes the four tail-end bins with coherent residuals, $r_s$ and $V_c(R_0)$ are fixed, and the density profile is not included in the dynamical likelihood. The posterior disfavors a strongly prolate force field in this restricted convention and prefers an oblate solution for the inner halo potential parameter. Other stream analyses show that oblate inner halo inferences are plausible: \citet{Malhan2019} obtained a GD-1 halo density flattening $q_\rho=0.82^{+0.25}_{-0.13}$ from Gaia, SEGUE, and LAMOST constraints; a global Gaia DR3 stream analysis by \citet{Ibata2024} favored a flattened halo with $q_{m,h}=0.75\pm0.03$; and the direct GD-1 acceleration analysis of \citet{NibauerBonaca2025} found $q_{\rho,z}=0.81^{+0.06}_{-0.03}$ together with evidence for a tilted triaxial halo.

\subsection{Modeling limitations}
\label{subsec:limitations}

The statistical uncertainty in Equation~\ref{eq:qresult} is conditional on the fixed bulge+disk+halo potential family, fixed $r_s=20\,{\rm kpc}$, fixed $V_c(R_0)=232\,\kms$, the potential-flattened NFW convention, the simplified particle release stream model with fixed progenitor mass and scale, and the adopted \Pal\ phase-space priors. Relaxing any of these assumptions can broaden or shift the posterior. The mild displacement of the posterior proper motions from the external priors indicates residual tension between the photometric track and the adopted static model.

The mass and scale tests isolate the response to fixed progenitor assumptions. The real cluster loses mass and changes its internal structure during disruption, so stars released at different times need not have the same release offsets or velocity distribution. This time evolution is not represented by a fixed Plummer model and contributes additional modeling uncertainty. Joint constraints on the remnant structure and the stream's kinematics would help distinguish this evolution from changes in the Galactic potential.

The dynamical likelihood uses the density profile only indirectly through the S/N selection of accepted bins. A density likelihood would add information on gaps, overdensities, and asymmetric stripping, but it requires the artificial stream response calibration discussed above. The next modeling stage should combine the DR10 star count morphology with Gaia proper motions, radial velocities, RR Lyrae and other distance tracers, and a density likelihood based on an explicit selection function. A model including a rotating bar, disk disequilibrium, globular cluster fly-bys, the LMC, and possible dark subhalo perturbations would then test whether the same DR10 track can be fit with a different effective halo flattening.

\section{Conclusions}
\label{sec:summary}

We have used DR10 $g,r,z$ photometry to measure the wide-field morphology of the \Pal\ stellar stream and to interpret its centroid and width with forward-modeled mock streams. The photometric selection combines a color--color locus, a cluster-calibrated PARSEC CMD filter, a continuous MSTO+RR Lyrae distance relation, and an empirical field star response tied to local $g$-band depth. This selection yields a homogeneous DR10 reference morphology over a substantially larger footprint than the high-confidence stream region.

The significant response-corrected stream spans $-15.5^\circ\le\phione\le7.0^\circ$. The leading arm remains broader and lower contrast than the inner trailing arm, consistent with previous evidence for leading arm fanning. The expanded search footprint improves the control of depth and foreground gradients and reveals no coherent additional low-surface-brightness continuation of the stream at the current selection threshold. This non-detection is best interpreted as a detection boundary measurement.

After masking the four coherent tail-end residuals, the fiducial 26-bin mock stream fit gives $\qphiz=0.824^{+0.133}_{-0.110}$ for the NFW potential flattening parameter when $r_s=20\,{\rm kpc}$ and $V_c(R_0)=232\,\kms$ are fixed. The corresponding total potential force diagnostic at the \Pal\ position is $q_{\Phi,{\rm total}}=0.862^{+0.064}_{-0.053}$. The fiducial posterior favors an oblate effective inner halo potential in the adopted static bulge+disk+halo model. Varying the fixed effective progenitor mass from $12,200$ to $47,000\,\msun$ at a 15 pc scale changes the median $\qphiz$ from $0.767$ to $0.889$, whereas changing the scale from 4 to 15 pc at $20,000\,\msun$ shifts the median by $-0.019$.

The present dynamical constraint is limited by systematic uncertainties in the potential family, progenitor mass and structural evolution, the phase-space priors, the tail-end mask, the lack of a density likelihood, and time-dependent perturbations from the Galactic bar, disk disequilibrium, the LMC, globular cluster fly-bys, and dark substructure. Because the bulge and disk parameters, the NFW scale radius, the circular-speed normalization, and the progenitor Plummer model are fixed within each fit, we do not independently determine the halo radial mass profile, the halo density axis ratio, the uncertainties in the baryonic potential, or the present-day mass, size, and mass-loss history of the \Pal\ progenitor. The axisymmetric model also does not constrain halo triaxiality or its orientation. In addition, a centroid-and-width likelihood alone cannot determine the physical origin of the coherent tail-end residuals or infer the masses and encounter geometries of possible perturbers. These limitations can be reduced by combining deeper multi-epoch imaging with improved astrometry and wide-field spectroscopy. The Vera C. Rubin Observatory Legacy Survey of Space and Time will extend the member sample to fainter main-sequence stars and provide a deeper and more homogeneous map of the low-surface-brightness tails. Its repeated imaging will also enable proper-motion measurements for stars fainter than the Gaia limit \citep{ivezic2019lsst}. Longer-baseline Gaia astrometry will improve bright-member selection and measurements of the proper-motion track, while targeted observations with the DESI Milky Way Survey and other spectroscopic surveys can provide line-of-sight velocities, metallicities, and chemical-abundance information \citep{cooper2023overview}. These measurements will provide stronger constraints on the progenitor's internal dispersion and stripping history.

\section*{Acknowledgments}
\noindent The authors acknowledge support from the National Key R\&D Program of China (grant Nos. 2023YFA1608100), the National Natural Science Foundation of China (NSFC; grant Nos. 12120101003, 12373010, 12233008, and 12503019) and China Manned Space Project (No. CMS-CSST-2025-A06). The authors also acknowledge the National Key R\&D Program of China (Grant Nos. 2023YFA1607804, 2022YFA1602902, and 2025YFE0202300), the Strategic Priority Research Program of the Chinese Academy of Sciences with Grant Nos. XDB0550100 and XDB0550000 and the Programs of National Astronomical Observatories Chinese Academy of Sciences with Grant Nos. E5ZQ7801, E5ZB7801, and E4TG2001.

\noindent This paper uses data obtained by the Legacy Surveys. The Legacy Surveys consist of three individual and complementary projects: the Dark Energy Camera Legacy Survey (DECaLS; Proposal ID 2014B-0404; PIs David Schlegel and Arjun Dey), the Beijing-Arizona Sky Survey (BASS; NOAO Proposal ID 2015A-0801; PIs Zhou Xu and Xiaohui Fan), and the Mayall $z$-band Legacy Survey (MzLS; Proposal ID 2016A-0453; PI Arjun Dey). DECaLS, BASS, and MzLS include data obtained, respectively, at the Blanco telescope, Cerro Tololo Inter-American Observatory, NSF's NOIRLab; the Bok telescope, Steward Observatory, University of Arizona; and the Mayall telescope, Kitt Peak National Observatory, NOIRLab. Pipeline processing and analyses of the data were supported by NOIRLab and Lawrence Berkeley National Laboratory. The Legacy Surveys project is honored to be permitted to conduct astronomical research on Iolkam Du'ag (Kitt Peak), a mountain with particular significance to the Tohono O'odham Nation. NOIRLab is operated by the Association of Universities for Research in Astronomy under a cooperative agreement with the National Science Foundation. Lawrence Berkeley National Laboratory is managed by the Regents of the University of California under contract to the U.S. Department of Energy.

\software{Astropy \citep{Astropy2013,Astropy2018,Astropy2022}, NumPy \citep{Harris2020NumPy}, SciPy \citep{Virtanen2020}, Matplotlib \citep{Hunter2007Matplotlib}, gala \citep{PriceWhelan2017Gala}, emcee \citep{ForemanMackey2013}}

\appendix

\section{Palomar 5 stream coordinate system}
\label{app:coords}

We use the heliocentric \Pal\ stream frame implemented in \texttt{gala.coordinates.Pal5PriceWhelan18}. For an International Celestial Reference System (ICRS) unit vector
\[
\mathbf{r}_{\rm ICRS}=\begin{bmatrix} \cos\alpha\cos\delta\\ \sin\alpha\cos\delta\\ \sin\delta
\end{bmatrix},
\]
the corresponding stream frame unit vector is
\[
\mathbf{r}_{\rm Pal5}=\begin{bmatrix} \cos\phione\cos\phitwo\\ \sin\phione\cos\phitwo\\ \sin\phitwo
\end{bmatrix}
=\mathbf{R}_{\rm Pal5}\,\mathbf{r}_{\rm ICRS},
\]
with
\[
\mathbf{R}_{\rm Pal5}=\begin{bmatrix}
-0.65582036 & -0.75491389 & -0.00216422\\
-0.62865701 & \phantom{-}0.54771927 & -0.55208149\\
\phantom{-}0.41795937 & -0.36070573 & -0.83378735
\end{bmatrix}.
\]
The longitude $\phione$ is wrapped at $180^\circ$, and the additional rotation in the \texttt{gala} implementation places the \Pal\ cluster near $(\phione,\phitwo)=(0,0)$. All trigonometric arguments in the equations above are angles in a common angular unit.

\section{Local morphology fits}
\label{app:localfits}

\begin{figure*}[htbp]
\centering
\includegraphics[width=0.92\textwidth]{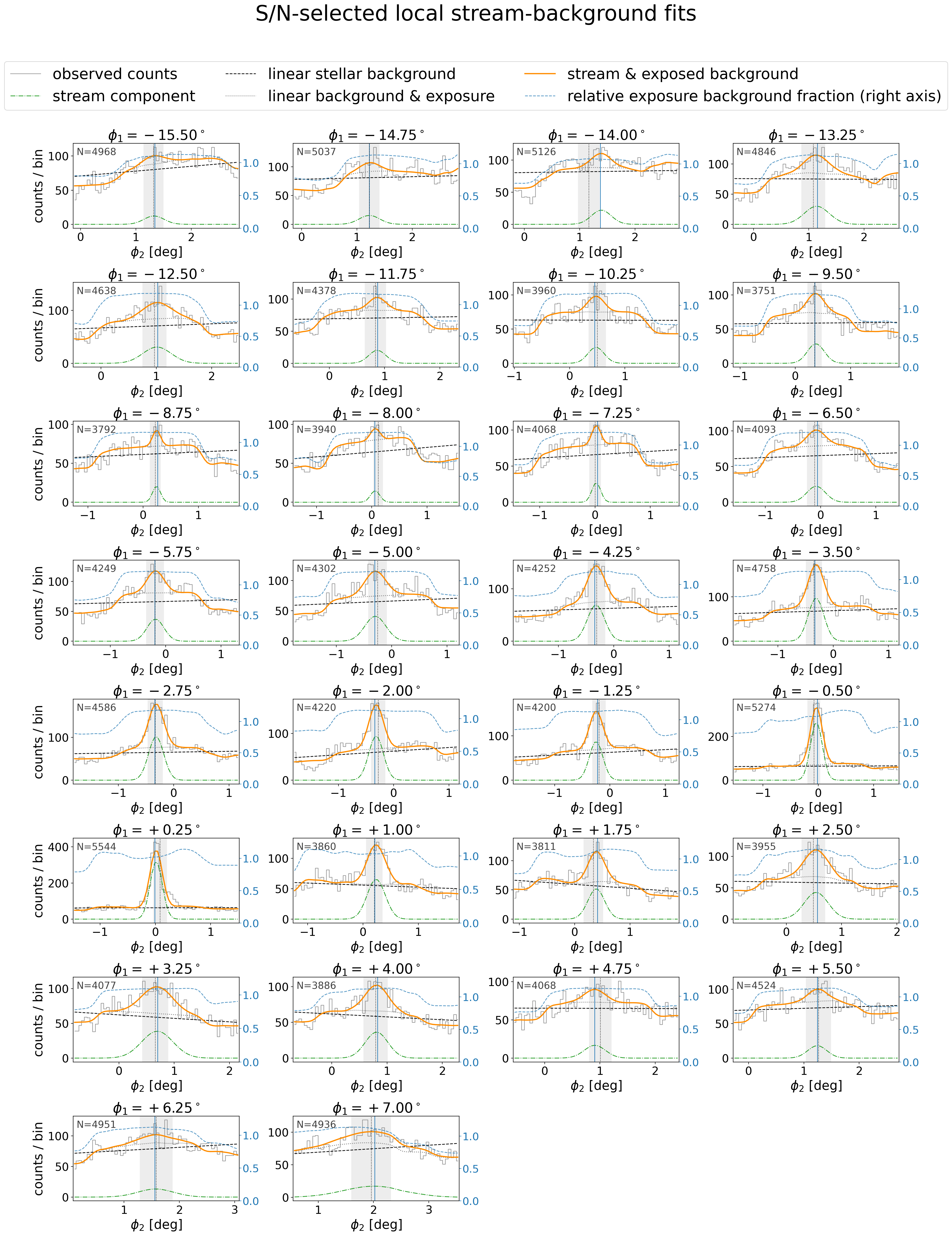}
\caption{Local stream-plus-background fits for the 30 $\phione$ bins that pass the stream count S/N threshold. Each panel shows the cross-stream selected star profile in one $\phione$ bin, together with the fitted total intensity, stream component, and local background component. The displayed fits define the measured centroid, width, stream count, and stream count S/N.}
\label{fig:localfits}
\end{figure*}

Figure~\ref{fig:localfits} shows the local stream-plus-background fits for the 30 bins that pass the stream count S/N criterion. Table~\ref{tab:morphtable} lists the 41 $\phione$ bins used to define the observed morphology, including S/N-rejected bins and the 26-bin subset used in the fiducial $q_z$ fit.

\clearpage

\startlongtable
\begin{deluxetable*}{rrrrrrrrrcc}
\tabletypesize{\scriptsize}
\tablecaption{Local morphology measurements for all $\phione$ bins. \label{tab:morphtable}}
\tablehead{
\colhead{$\phione$}(deg) & \colhead{$\mu$}(deg) & \colhead{$s_{\mu,-}$}(deg) & \colhead{$s_{\mu,+}$}(deg) & \colhead{$\sigma$}(deg) & \colhead{$s_{\sigma,-}$}(deg) & \colhead{$s_{\sigma,+}$}(deg) & \colhead{$N_{\rm str}$} & \colhead{S/N} & \colhead{S/N flag} & \colhead{$q_z$ flag}
}
\startdata
-20.00 & 1.986 & 0.120 & 0.125 & 0.599 & 0.083 & 0.092 & 0.1 & 0.03 & N & N \\
-19.25 & 1.956 & 0.118 & 0.125 & 0.341 & 0.048 & 0.055 & 0.1 & 0.05 & N & N \\
-18.50 & 1.816 & 0.122 & 0.106 & 0.248 & 0.034 & 0.041 & 3.3 & 0.05 & N & N \\
-17.75 & 1.764 & 0.107 & 0.096 & 0.193 & 0.028 & 0.031 & 3.3 & 0.05 & N & N \\
-17.00 & 1.549 & 0.117 & 0.118 & 0.167 & 0.023 & 0.027 & 0.1 & 0.04 & N & N \\
-16.25 & 1.415 & 0.112 & 0.101 & 0.171 & 0.025 & 0.029 & 15.4 & 0.32 & N & N \\
-15.50 & 1.327 & 0.067 & 0.072 & 0.180 & 0.024 & 0.029 & 110.7 & 1.79 & Y & N \\
-14.75 & 1.225 & 0.102 & 0.083 & 0.186 & 0.031 & 0.037 & 143.0 & 2.39 & Y & N \\
-14.00 & 1.392 & 0.044 & 0.046 & 0.178 & 0.027 & 0.035 & 196.9 & 4.01 & Y & Y \\
-13.25 & 1.146 & 0.043 & 0.045 & 0.226 & 0.038 & 0.043 & 338.5 & 5.03 & Y & Y \\
-12.50 & 1.012 & 0.040 & 0.038 & 0.257 & 0.032 & 0.034 & 393.5 & 5.95 & Y & Y \\
-11.75 & 0.870 & 0.042 & 0.042 & 0.155 & 0.019 & 0.021 & 157.0 & 3.64 & Y & Y \\
-11.00 & 0.623 & 0.095 & 0.094 & 0.196 & 0.028 & 0.033 & 8.3 & 0.12 & N & N \\
-10.25 & 0.478 & 0.034 & 0.032 & 0.147 & 0.016 & 0.018 & 172.4 & 4.49 & Y & Y \\
-9.50 & 0.373 & 0.032 & 0.030 & 0.153 & 0.018 & 0.019 & 219.3 & 5.16 & Y & Y \\
-8.75 & 0.241 & 0.027 & 0.026 & 0.072 & 0.009 & 0.010 & 74.5 & 3.13 & Y & Y \\
-8.00 & 0.061 & 0.037 & 0.046 & 0.084 & 0.011 & 0.013 & 62.7 & 1.80 & Y & Y \\
-7.25 & 0.023 & 0.023 & 0.022 & 0.080 & 0.010 & 0.011 & 109.2 & 3.76 & Y & Y \\
-6.50 & -0.081 & 0.044 & 0.043 & 0.167 & 0.021 & 0.024 & 187.5 & 4.08 & Y & Y \\
-5.75 & -0.184 & 0.024 & 0.026 & 0.153 & 0.018 & 0.020 & 284.1 & 6.02 & Y & Y \\
-5.00 & -0.294 & 0.025 & 0.025 & 0.190 & 0.020 & 0.024 & 392.1 & 7.93 & Y & Y \\
-4.25 & -0.322 & 0.015 & 0.014 & 0.157 & 0.013 & 0.015 & 542.2 & 13.04 & Y & Y \\
-3.50 & -0.309 & 0.011 & 0.010 & 0.141 & 0.008 & 0.009 & 688.4 & 13.54 & Y & Y \\
-2.75 & -0.317 & 0.010 & 0.009 & 0.137 & 0.008 & 0.009 & 699.2 & 15.08 & Y & Y \\
-2.00 & -0.318 & 0.010 & 0.010 & 0.132 & 0.008 & 0.009 & 626.6 & 12.05 & Y & Y \\
-1.25 & -0.251 & 0.010 & 0.010 & 0.121 & 0.008 & 0.009 & 531.7 & 10.28 & Y & Y \\
-0.50 & -0.035 & 0.004 & 0.004 & 0.105 & 0.004 & 0.005 & 1427.1 & 26.24 & Y & Y \\
0.25 & 0.014 & 0.004 & 0.004 & 0.110 & 0.005 & 0.005 & 1773.0 & 27.03 & Y & Y \\
1.00 & 0.236 & 0.014 & 0.014 & 0.163 & 0.012 & 0.013 & 538.9 & 11.39 & Y & Y \\
1.75 & 0.385 & 0.018 & 0.017 & 0.153 & 0.015 & 0.017 & 395.9 & 9.01 & Y & Y \\
2.50 & 0.541 & 0.027 & 0.027 & 0.219 & 0.026 & 0.028 & 462.3 & 7.90 & Y & Y \\
3.25 & 0.694 & 0.034 & 0.031 & 0.282 & 0.026 & 0.030 & 547.3 & 6.80 & Y & Y \\
4.00 & 0.810 & 0.026 & 0.027 & 0.205 & 0.021 & 0.022 & 373.3 & 7.08 & Y & Y \\
4.75 & 0.909 & 0.050 & 0.052 & 0.192 & 0.025 & 0.028 & 161.0 & 3.35 & Y & Y \\
5.50 & 1.242 & 0.049 & 0.046 & 0.172 & 0.021 & 0.025 & 155.6 & 3.77 & Y & Y \\
6.25 & 1.575 & 0.072 & 0.073 & 0.329 & 0.040 & 0.049 & 213.8 & 3.12 & Y & N \\
7.00 & 2.046 & 0.088 & 0.092 & 0.581 & 0.077 & 0.090 & 465.4 & 3.11 & Y & N \\
7.75 & 2.318 & 0.116 & 0.115 & 0.626 & 0.083 & 0.087 & 0.7 & 0.02 & N & N \\
8.50 & 2.727 & 0.119 & 0.113 & 0.676 & 0.081 & 0.075 & 0.2 & 0.02 & N & N \\
9.25 & 3.150 & 0.120 & 0.117 & 0.621 & 0.085 & 0.090 & 1.1 & 0.02 & N & N \\
10.00 & 3.490 & 0.119 & 0.121 & 0.621 & 0.082 & 0.088 & 0.2 & 0.01 & N & N \\
\enddata
\tablecomments{The S/N flag indicates whether a bin satisfies ${\rm S/N}_{\rm stream}>1$, and the $q_z$ flag indicates whether it is included in the fiducial dynamical fit. $N_{\rm str}$ is the expected number of selected stream stars in the local window.}
\end{deluxetable*}
\clearpage

\bibliographystyle{aasjournal}
\bibliography{references}

\end{document}